\documentclass[a4paper,10pt]{article}
\usepackage[a4paper, margin=2cm]{geometry}

\usepackage[utf8]{inputenc}
\usepackage[T1]{fontenc}

\usepackage{fourier}

\usepackage[square,numbers]{natbib}
\usepackage{amsmath}
\allowdisplaybreaks
\usepackage{amsfonts}
\usepackage{bm}
\usepackage{mathrsfs}
\usepackage{booktabs}
\usepackage{graphicx}
\usepackage{url}
\usepackage{xcolor}

\usepackage{amsthm}
\theoremstyle{remark}
\newtheorem*{remark}{Remark}

\usepackage[english]{babel}
\usepackage[hidelinks]{hyperref}

\usepackage{doi}

\usepackage{ulem} 

\title{Elastic wave propagation in magneto-active fibre composites}
\author{Harold Berjamin \textsuperscript{a}, Stephan Rudykh \textsuperscript{a} \\ ~ \\
\emph{\footnotesize\textsuperscript{a}School of Mathematical and Statistical Sciences, University of Galway, University Road, Galway, Republic of Ireland}}

\date{}

\begin{document}

	\maketitle
	
	\begin{abstract}
		\noindent
		Fibre-reinforced elastomers are lightweight and strong materials that can sustain large deformations. When filled with magnetic particles, their effective mechanical response can be modified by an external magnetic field. In the present study, we propose an effective theory of fibre-reinforced composite, based on a neo-Hookean elastic response and a linear magnetic law in each phase. The theory is shown suitable to describe the motion of composite cylinders. Furthermore, it is found appropriate for the modelling of fibre-reinforced composites subjected to a permanent magnetic field aligned with the fibres. To reach this result, we use the incremental theory (`small on large'), in combination with homogenisation theory and the Bloch-Floquet method. This way, we show that wave directivity is sensitive to the application of a permanent magnetic field, whereas the frequency range in which wave propagation is forbidden is not modified by such a load (the band gaps are invariant). In passing, we describe a method to deduce the total stress in the material based on the measurement of two wave speeds. Furthermore, we propose an effective energy function for the description of nonlinear composites made of Yeoh-type generalised neo-Hookean fibres within a neo-Hookean matrix.
		~ \\
		\emph{Keywords:~} Magnetoactive elastomers, Fibre-reinforced composites, Finite deformations, Shear waves, Phononic crystals
	\end{abstract}

\section{Introduction}\label{sec:Intro}

Magneto-active elastomers are soft rubber-like materials filled with magnetic particles. Thus, by construction, their mechanical response is coupled to the magnetic response of the particles. For instance, under the application of an external magnetic field, the shape of such a material sample can be modified in a variety of ways \citep{bastola21}. Therefore, these materials are of great interest for potential applications in robotics \citep{kim22}, among other fields. For a review of recent applications in 3D printing, soft robotics and sensor/actuator physics, the interested reader is referred to \citet{khalid24}.

Soft magneto-active solids can be separated into two main categories according to the type of magnetic particles that were inserted in the rubber-like solid, see for instance the review by \citet{saber23}. On the one hand, \emph{soft} magnetic particles do not remain magnetised in the absence of an external magnetic field, by opposition to \emph{hard} magnetic particles which exhibit a residual magnetisation, on the other hand. In the present study, we restrict our attention to the former class of materials, which is characterised by a low residual magnetisation and a low magnetic coercivity.

Various constitutive models for the analysis of soft magneto-active elastomers have been proposed over the years, with contributions by \citet{dorfmann04} as well as \citet{kankanala04}. In these works, the coupling between mechanical and magnetic forces is described, and a finite-strain continuum theory for the modelling of isotropic magneto-active materials is introduced. The proposed theory is supported by homogenisation theory, see the works by \citet{castaneda11} in which the macroscopic response is linked to the properties of the elastomer and of the particles. Among other applications, this theory was used to investigate the propagation of waves in isotropic magneto-active materials \citep{destrade11}, the effect of the particle volume fraction in silicone polymer samples \citep{garciagonzalez21}, as well as the wrinkling instability in magneto-active plates \citep{wu21}.

Based on these earlier works, the modelling of transversely isotropic (TI) soft magneto-active elastomers was addressed by \citet{bustamante07}, see also \citet{bustamante10}. Contrary to the isotropic case, TI materials have a distinct response along a preferred direction, whose orientation is described by a material vector. For instance, such materials can be manufactured by alignment of the magnetic particles along chain-like structures with a given direction in the rubber-like matrix \citep{danas12, saxena15}. Alternatively, a TI magneto-active material can be manufactured by lamination, that is by superposition of multiple layers of isotropic magneto-active elastomers \citep{rudykh13}.

Following the latter approach, we investigate the magneto-mechanical response of composite materials made of magneto-active fibres embedded within a magneto-active matrix. Our contribution is the development of an effective theory for fibre-reinforced composites, which belongs to the class of TI magneto-active elastomers. Here, we assume the material incompressible with neo-Hookean behaviour, and the magnetic law is assumed linear. We then investigate the dynamic properties of this microstructured medium, in a similar fashion to \citet{karami19} who considered magneto-active laminates. Here, the results deduced from the proposed effective theory are compared to numerical results obtained directly from the heterogeneous structure. Doing so, we find that wave directivity is sensitive to the application of a permanent magnetic field, whereas the frequency range in which wave propagation is forbidden is not modified by such a load (the \emph{band gaps} are invariant).

Our study is also related to that by \citet{zhang22}, as well as \citet{alam23, padmanabhan24}, where the dynamic response of 1D and 2D hard-magnetic phononic crystals is discussed. In these works, a model of hard magnetic elastomer is used to establish the controllability of the frequency range in which wave propagation is forbidden (here, band gaps are tunable). It is worth pointing out in passing that the underlying theory should be updated to account for recent improvements, see \citet{dorfmann24}. For a unified theory, the interested reader is referred to relevant works on the modelling of magneto-active solids \citep{morro24}, see also \citet{lucarini22} for a literature review on hard-magnetic materials.

The paper is organised as follows. In Section~\ref{sec:Constitutive}, the equations governing soft magneto-deformations are presented, and the TI theory \eqref{EnergyTI} describing fibre-reinforced composites is introduced. In Section~\ref{sec:Shear}, we carry out the derivation of this theory using the out-of-plane shear and extension deformation of a composite cylinder. In Section~\ref{sec:SoL}, we then investigate the incremental response of this material by considering a small perturbation of a large static magneto-deformation (`small on large'). We use these results to analyse the connection between the static stress and the speed of shear waves in the material. In Section~\ref{sec:2D}, we combine the small-on-large theory with periodic homogenisation as well as with the Floquet-Bloch technique to study shear wave propagation in pre-stressed composites. Concluding remarks can be found in Section~\ref{sec:Conclu}. The appendices \ref{app:OtherConst}-\ref{app:Torsion} discuss potential generalisations of the present results, including a study of the effective response of a nonlinear composite made of Yeoh fibres within a neo-Hookean matrix, for which an effective energy function is derived (Appendix~\ref{app:I1}).

\section{Constitutive theory}\label{sec:Constitutive}

In the present study, we consider an extension of the linear theory for incompressible magneto-elastic solids with a neo-Hookean behaviour and a linear magnetic response. This situation might represent rigid, magnetically isotropic, and magnetically-active spherical particles, embedded in a neo-Hookean elastomeric matrix \citep{castaneda11}. The isotropic theory is then used to derive a transversely isotropic (TI) theory that describes the motion of a given composite material based on two isotropic phases.

Let us introduce the deformation gradient tensor $\bm{F} = \partial{\bm x}/\partial{\bm X}$, which is the gradient of the map $(\bm{X},t) \mapsto \bm{x}$ that connects the position ${\bm x}$ of a material point in the current configuration at time $t$ to its position $\bm X$ in the reference configuration. In soft incompressible solids, the determinant of the deformation gradient tensor is prescribed, i.e., the constraint $\det \bm{F} = 1$ is enforced. To describe solids that are initially isotropic, one may introduce the total magneto-elastic energy $\Psi$ which is defined in terms of five scalar invariants \citep{dorfmann04}. In general, these invariants involve the magnetic flux density vector in the reference configuration, ${\bf b}_0$ (or Lagrangian magnetic induction, expressed in T), and a Lagrangian strain tensor, such as the right Cauchy-Green strain tensor $\bm{C} = \bm{F}^\text{T}\bm{F}$. The total stress tensor $\bm{\sigma}$ and Lagrangian magnetic field strength vector ${\bf h}_0$ (in A/m) are deduced from
\begin{equation}
    \bm{\sigma} = \bm{\sigma}^\text{E} - p{\bm I}, \quad
    \bm{\sigma}^\text{E} = \frac{\partial \Psi}{\partial \bm F}\bm{F}^\text{T} , \quad 
    {\bf h}_0 = \frac{\partial \Psi}{\partial {\bf b}_0} .
    \label{ConstitutiveDer}
\end{equation}
Here, the notation $\bm I$ stands for the second-order identity tensor, the exponent \textsuperscript{T} signifies the transpose, $\bm{\sigma}^\text{E}$ is the extra stress, and $p$ (in Pa) is a Lagrange multiplier to enforce incompressibility.

In the absence of external body forces, the mechanical equilibrium equations take the form
\begin{equation}
    \text{div}\, \bm{\sigma} = \bm{0} , \quad \bm{\sigma} = \bm{\sigma}^\text{T} ,
    \label{Equil}
\end{equation}
where $\text{div}$ is the divergence operator in the current configuration.
Furthermore, in the absence of a prescribed current density, the Maxwell field equations read
\begin{equation}
    \begin{aligned}
    &\text{curl}\, {\bf h} = {\bf 0}, &\quad {\bf h} &= \bm{F}^{-\text{T}} {\bf h}_0 ,
    \\
    &\text{div}\, {\bf b} = 0, &\quad {\bf b} &= \bm{F} {\bf b}_0 ,
    \end{aligned}
    \label{Maxwell}
\end{equation}
which are Ampère's and Gauss' laws of magnetostatics, respectively. Here, the vectors ${\bf b}$ and ${\bf h}$ are the magnetic induction and magnetic strength fields in the current configuration, that is, they are the Eulerian counterparts of ${\bf b}_0$ and ${\bf h}_0$.
At an interface with unit normal $\hat{\bf n}$ between two such materials, the normal tractions $\bm{\sigma} \cdot \hat{\bf n}$ must be continuous, as well as the quantities $\bf{h} \times \hat{\bf n}$ and $\bf{b}\cdot \hat{\bf n}$ when no surface current density is prescribed at the boundary.

In what follows, we consider a class of transversely isotropic (TI) magneto-elastic solids defined by the total energy function
\begin{equation}
    \Psi = \frac{\tilde G}2 (I_1 - 3) + \frac{\bar G - \tilde G}2 \left(I_7 + 2 I_7^{-1/2} - 3\right) + \frac{I_5}{2\bar\mu} .
    \label{EnergyTI}
\end{equation}
The above expression involves the scalar invariants \citep{bustamante10}
\begin{equation}
    I_1 = \text{tr}(\bm{C}), \quad I_5 = {\bf b}_0\cdot \bm{C} {\bf b}_0, \quad I_7 = \hat{\bf n}_0 \cdot \bm{C} \hat{\bf n}_0,
    \label{Invar}
\end{equation}
where $\hat{\bf n}_0$ is a unit vector along the direction of anisotropy in the reference configuration. The parameter $\bar G$ is the isochoric extension modulus and $\tilde G$ represents the in-plane and out-of-plane shear modulus. The parameter $\bar\mu$ (in N/A\textsuperscript{2}) is the magnetic permeability. Depending on conventions, the (dimensionless) magnetic permeability is sometimes defined relatively to the permeability of vacuum $\mu_0 = 4\pi \times 10^{-7}$ N/A\textsuperscript{2}, see for instance \citet{dorfmann04}.

Formulas for $\tilde G$, $\bar G$, and $\bar \mu$ that correspond to a fibre-reinforced material will be established in Section~\ref{sec:Shear}. Therein, we consider cylindrical fibres with shear modulus $\tilde G = \bar G = G^\text{(f)}$ and magnetic permeability $\bar \mu = \mu^\text{(f)}$ embedded within a similar host material with parameters $G^\text{(s)}$, $\mu^\text{(s)}$. Both materials are assumed isotropic and described by the energy function \eqref{EnergyTI} with two parameters only, the shear modulus and the magnetic permeability. Under the assumption of perfect bonding, we have
\begin{equation}
    \begin{aligned}
    \bar{G} &= n^\text{(f)} G^\text{(f)} + n^\text{(s)} G^\text{(s)}, \\ 
    \tilde{G} &= n^\text{(f)}\alpha^\text{(f)} G^\text{(f)} + n^\text{(s)}\alpha^\text{(s)} G^\text{(s)} , \\
    \bar{\mu}\, &= n^\text{(f)} \mu^\text{(f)} + n^\text{(s)} \mu^\text{(s)} ,
    \end{aligned} 
    \label{Effective}
\end{equation}
where
\begin{equation}
    \begin{aligned}
    \alpha^\text{(f)} &= \frac{2 G^\text{(s)}}{n^\text{(s)}G^\text{(f)} + (1+n^\text{(f)})G^\text{(s)}} ,\\
    \alpha^\text{(s)} &= \frac{G^\text{(f)} + G^\text{(s)}}{n^\text{(s)}G^\text{(f)} + (1+n^\text{(f)})G^\text{(s)}},
    \end{aligned}
    \label{deBotton}
\end{equation}
together with the relationship $n^\text{(f)}\alpha^\text{(f)} + n^\text{(s)}\alpha^\text{(s)} = 1$. Here, the quantity $n^\text{(f)} = 1 - n^\text{(s)}$ represents the volume fraction of fibre material, and $0 \leq n^\text{(s)} \leq 1$ is the volume fraction of host material.

With the expression \eqref{EnergyTI} of the total energy, we arrive at the following expression of the extra stress \eqref{ConstitutiveDer}
\begin{equation}
    \bm{\sigma}^\text{E} = 2 \Psi_1 {\bm B} + 2\Psi_7 {\bf n}\otimes {\bf n} + {\bf h}\otimes{\bf b} , \quad {\bf b} = {\bf h}/(2\Psi_5) ,
    \label{ConstitutiveExpr}
\end{equation}
where $\bm{B} = \bm{F}\bm{F}^\text{T}$ is the left Cauchy-Green strain tensor. Here, we have introduced the coefficients $\Psi_n = \partial \Psi / \partial I_n$ obtained by differentiation of $\Psi$ with respect to the invariants \eqref{Invar},
\begin{equation}
    \Psi_1 = \frac{\tilde G}{2} , \quad
    \Psi_5 = \frac1{2\bar\mu} , \quad
    \Psi_7  = \frac{\bar G - \tilde G}2 \left(1- I_7^{-3/2}\right) .
    \label{Psi7Coeff}
\end{equation}
The vector ${\bf n} = \bm{F} \hat{\bf n}_0$ marks the direction of anisotropy in the deformed configuration (note that $\|{\bf n}\|^2 = I_7$).
In vacuum, the total stress arising from the magnetic field is the Maxwell stress
\begin{equation}
    \bm{\sigma}^\text{M} = {\bf h}\otimes {\bf b} - \tfrac12({\bf h}\cdot{\bf b})\bm{I} , \quad {\bf b} = \mu_0 {\bf h},
    \label{StressMax}
\end{equation}
to be used for the boundary conditions at a free interface. The magnetic permeability of vacuum $\mu_0$ was introduced earlier.

For the above energy function, the dependency of $\Psi$ with respect to the invariant $I_5 = \|{\bf b}\|^2$ follows from the linear magneto-elastic theory \citep{dorfmann04}. Within this framework, the definition of the magnetisation ${\bf m} = {\bf b}/\mu_0 - {\bf h}$ (in A/m) leads to
\begin{equation}
    {\bf m} = \frac{\bar\mu - \mu_0}{\mu_0} {\bf h} = \frac{\bar\mu-\mu_0}{\bar\mu \mu_0} {\bf b} ,
    \label{Magne}
\end{equation}
where we have used the relationship ${\bf b} = \bar{\mu} {\bf h}$ introduced in Eqs.~\eqref{ConstitutiveExpr}-\eqref{Psi7Coeff}. Therefore, the magnetic vector fields ${\bf m}$, ${\bf h}$, ${\bf b}$ are always aligned in such a material. Other forms of the magneto-elastic energy \eqref{EnergyTI} could be considered to account for the magnetic saturation effect that arises at large values of $\|{\bf b}\|$, see for instance the Langevin function \citep{wu21}.

In this theory described by the energy function \eqref{EnergyTI} and the extra stress \eqref{ConstitutiveExpr}, the term proportional to $(\bar G - \tilde G)$ accounts for the anisotropic mechanical response of the composite material. For neo-Hookean elastic solids, this term is included in an exact model for the out-of-plane shear and extensional motion of a composite cylinder \citep{debotton06b}. The same expression can also be used to model the homogenous deformations of composite materials with neo-Hookean phases \citep{guo07}. Other forms of the strain energy function are found in the literature to model solids reinforced by polydisperse cylindrical fibres of infinitely many diameters, cf. \citet{lopezjimenez14}.

The present theory recovers the \emph{standard reinforcing model} in the limit of small deformations $I_7 \simeq 1$, for which \eqref{Psi7Coeff} entails $2\Psi_7 \simeq E (I_7-1)$, and $E = \frac32 (\bar G - \tilde G)$ represents the degree of anisotropy, see \citet{horgan12}. The ratio $E/(2\tilde G)$ is the reinforcing
strength parameter in the fibre direction \citep{merodio01}. By studying the evolution of the quantity $E / G^\text{(f)}$ with respect to the variables $n^\text{(f)} \in [0, 1]$ and $G^\text{(s)}/G^\text{(f)} \geq 0$, see expressions in Eq.~\eqref{Effective}, it can be shown that $E \geq 0$ for all volume fractions of fibre and for all shear stiffness ratios. For later use, we define the scalar $q$ such that
\begin{equation}
    p = q - (\bar G - \tilde G) I_7^{-1/2} ,
    \label{SubsBotton}
\end{equation}
which is a suitable redefinition of the arbitrary Lagrange multiplier introduced to enforce incompressibility in Eq.~\eqref{ConstitutiveDer}.

 \section{Out-of-plane shear and extension of a magneto-elastic composite cylinder}\label{sec:Shear}

In this section, we apply the above theory to the case of a cylinder subjected to an out-of-plane shearing and extensional motion defined subsequently. We consider a solid cylinder made of this TI material \eqref{EnergyTI}, with $\hat{\bf n}_0$ oriented vertically, along the cylinder's axis. The cylinder has infinite length and radius $A$ in the reference configuration, and deformed radius $a = A \lambda'$ where $\lambda'$ is the radial stretch. The motion is described by the cylindrical coordinates $(R, \Theta, Z)$ of a material point in the reference configuration, and by the coordinates $(r, \theta, z)$ in the deformed configuration.

\begin{figure}
    \centering
    \includegraphics{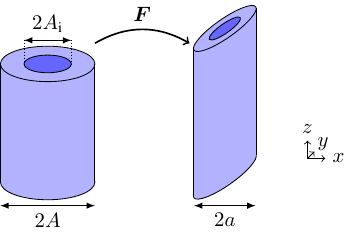}
    \caption{Section of a soft magneto-elastic composite cylinder subjected to out-of-plane shear and extension, with internal radius $A_\text{i}$ and outer radius $A$ before deformation (left), and outer radius $a = A \lambda'$ after deformation (right). For the homogeneous cylinder, the picture is very similar, but with a single material. \label{fig:Struct}}
\end{figure}

To enable direct comparisons, the same out-of-plane shear problem is then studied for a composite cylinder made of a fibre embedded inside a shell. Here, we will consider a perfectly bonded cylinder made of two isotropic neo-Hookean elastomers for which $\tilde G = \bar G$ in each phase{\,---\,}cf. \citet{danishevskyy15} for a study of composites with imperfect interfaces. The fibre located in the region $0\leq R \leq A_\text{i}$ is a magneto-elastic material described by the energy function \eqref{EnergyTI} with $\tilde G = G^\text{(f)}$ and $\bar\mu = \mu^\text{(f)}$, whereas the shell located in the region $A_\text{i}\leq R \leq A$ has parameters $\tilde G = G^\text{(s)}$, $\bar\mu = \mu^\text{(s)}$, see the notations in Figure~\ref{fig:Struct}.

Alternative material models are considered in the Appendix~\ref{app:OtherConst}. More specifically, a generalisation \eqref{EnergyYeohTI} of the energy function \eqref{EnergyTI} is considered in the Appendix~\ref{app:I1}. Material models based on the second strain invariant $I_2$ are investigated in the Appendix~\ref{app:I2}, where it is shown that the present motion is much more difficult to analyse for this particular class of materials. Furthermore, the case of torsional motions with uniaxial extension is addressed in the Appendix~\ref{app:Torsion} based on the constitutive theory \eqref{EnergyTI}.

\subsection{The homogeneous TI cylinder}

First, we study the out-of-plane shear deformation and extension of an infinite homogeneous cylinder, which results from an imposed displacement at its lateral boundaries. The motion is described by \citep{debotton06b}
\begin{equation}
    r = \lambda' R, \quad \theta = \Theta, \quad z = \lambda Z + \gamma R \cos\Theta ,
    \label{Shear}
\end{equation}
where $\lambda = (\lambda')^{-2}$ is the axial stretch and $\gamma$ is the amount of shear. For this motion, the deformation gradient tensor relative to the present system of cylindrical coordinates is represented by the following matrix
\begin{equation}
    \bm{F} = \begin{bmatrix}
        \lambda' & 0 & 0\\
        0 & \lambda' & 0\\
        \gamma \cos\theta & -\gamma \sin\theta & \lambda
    \end{bmatrix} , \quad \lambda (\lambda^{\prime})^2 = 1,
    \label{FShear}
\end{equation}
which has unit determinant. For further details about finite deformations in cylindrical coordinates, see \citet{ogden84}.

The magnetic field $\bf h$ is assumed to be purely axial, i.e., ${\bf h} = h \hat{\bf e}_z$, where $\hat{\bf e}_z$ is the unit vector orientating the vertical axis. To satisfy the magnetostatic equations \eqref{Maxwell}, the axial component $h$ of the magnetic field needs to be constant, and the magnetic induction $b = \bar\mu h$ along $\hat{\bf e}_z$ is constant too. In the present case, the relevant scalar invariants \eqref{Invar} are given by
\begin{equation}
    I_1 = \lambda^2 + 2\lambda^{\prime 2} + \gamma^2, \quad I_5 = b^2, \quad I_7 = \lambda^2 .
    \label{InvarShear}
\end{equation}
Furthermore, in mechanical equilibrium \eqref{Equil}, the Lagrangian multiplier $p$ must be constant.

The non-zero tractions on the lateral boundary deduced from the constitutive law \eqref{ConstitutiveDer}-\eqref{ConstitutiveExpr} read
\begin{equation}
    \sigma_{rr} = \bar G\lambda^{\prime 2} - q, \quad \sigma_{rz} = \tilde G \lambda' \gamma \cos\theta ,
    \label{TShear}
\end{equation}
where the alternative Lagrange multiplier $q$ introduced in \eqref{SubsBotton} is a constant.
The magnetic field induces a modification of the tractions applied to the cylinder's cross section, more specifically through the stress component
\begin{equation}
    \begin{aligned}
    \sigma_{zz} &= \bar G \lambda^2 + \tilde G \gamma^2 + hb - q .
    \end{aligned}
    \label{NShear}
\end{equation}
Given that this traction is constant over the circular cross section, the resulting normal force is given by the formula
\begin{equation}
    \mathcal{N} = 2\pi \int_0^a r \sigma_{z z} \text{d}r = \pi \sigma_{zz} a^2 .
    \label{NormalForce}
\end{equation}

\subsection{The composite cylinder}

We study now the same problem for a composite cylinder, as shown in Figure~\ref{fig:Struct}. In this cylinder, the volume fraction of fibre $\varphi^\text{(f)} = (A_\text{i}/A)^2$ in the reference configuration satisfies $\varphi^\text{(f)} + \varphi^\text{(s)} = 1$, where $\varphi^\text{(s)}$ is the initial volume fraction of shell material. At its lateral boundary, the cylinder is subjected to a radial contraction of stretch ratio $\lambda' = \lambda^{-1/2}$, so that the cylinder radius satisfies $a = A\lambda'$. The volume fraction of fibre $n^\text{(f)} = (a_\text{i}/a)^2$ in the current configuration is equal to that in the reference configuration, $\varphi^\text{(f)}$, and a similar equality holds for the volume fractions of shell material. In what follows, the magnetic field ${\bf h} = h \hat{\bf e}_z$ is assumed uniform, and the magnetic induction $b^\bullet = \mu^\bullet h$ along $\hat{\bf e}_z$ is constant in each phase, marked by the bullet point exponents $\bullet \in \lbrace (\text{f}), (\text{s})\rbrace$.

In each phase, the cylinder is subjected to out-of-plane shear and extension defined by
\begin{equation}
    r = \lambda' R, \quad \theta = \Theta, \quad z = \lambda Z + \gamma f^\bullet(R) \cos\Theta ,
    \label{ShearComp}
\end{equation}
where the functions $f^\bullet$ are unknown. The deformation gradient reads
\begin{equation}
    \bm{F}^\bullet = \begin{bmatrix}
        \lambda' & 0 & 0\\
        0 & \lambda' & 0\\
        \gamma \frac{\text{d} f^\bullet}{\text{d} R} \cos\theta & -\gamma \frac{f^\bullet}{R} \sin\theta & \lambda
    \end{bmatrix} ,
    \label{FShearComp}
\end{equation}
which has unit determinant.
The relevant phase-related invariants \eqref{Invar} become
\begin{equation}
    \begin{aligned}
    &I_1^\bullet = \lambda^2 + 2\lambda^{\prime 2} + \gamma^2 \left( (\tfrac{\text{d} f^\bullet}{\text{d} R} \cos\theta)^2 + (\tfrac{f^\bullet}{R} \sin\theta)^2 \right) , \\
    &I_5^\bullet = (b^\bullet)^2 .
    \end{aligned}
    \label{InvarComp}
\end{equation}
Based on the assumption that the cylinder is infinite, the Lagrangian multiplier $p^\bullet$ is sought independent of $z$.

In a state of mechanical equilibrium \eqref{Equil}, the Lagrangian multiplier $p^\bullet$ must be constant in each phase. Following enforcement of the boundary conditions at the inner and outer interfaces of the cylinder, it can be shown that \citep{debotton06b}
\begin{equation}
    f^\bullet(R) = \alpha^\bullet R + \beta^\bullet A^2/ R ,
    \label{deBottonF}
\end{equation}
where the coefficients $\alpha^\bullet$ are given in Eq.~\eqref{deBotton} and
\begin{equation}
    \beta^\text{(f)} = 0 ,\quad
    \beta^\text{(s)} = \frac{(G^\text{(s)} - G^\text{(f)})\, n^\text{(f)}}{n^\text{(s)}G^\text{(f)} + (1+n^\text{(f)})G^\text{(s)}} = \alpha^\text{(s)} - \frac{\tilde G}{G^\text{(s)}} ,
    \label{deBotton2}
\end{equation}
see Eqs.~\eqref{Effective}-\eqref{deBotton} for the notations.
Furthermore, the pressure is subjected to a jump condition across the inner interface, $p^\text{(f)} - p^\text{(s)} = (G^\text{(f)}-G^\text{(s)}) \lambda^{\prime 2}$.

We can now evaluate the resulting tractions $\sigma_{r \star}$ applied to the lateral boundary ($\star = r, \theta, z$), which are equal to $\sigma^\text{(s)}_{r \star}|_{R=A}$. Similarly to \citet{debotton06b}, we arrive at the non-zero tractions
\begin{equation}
    \sigma_{rr} = \bar{G}\lambda^{\prime 2} - \bar{p} ,\quad
    \sigma_{r z} = \tilde{G}\lambda' \gamma  \cos\theta ,
    \label{EffTractions}
\end{equation}
with the effective quantities \eqref{Effective} and $\bar{p} = n^\text{(f)} p^\text{(f)} + n^\text{(s)} p^\text{(s)}$.
From the expressions \eqref{Effective}, it becomes obvious that the tractions \eqref{EffTractions} can be written in the form of a weighted sum of phase-related stress contributions with coefficients $n^\text{(f)}$, $n^\text{(s)}$.

Addition of a non-zero magnetic field modifies the vertical tractions, which are given by the formula
\begin{equation}
    \sigma_{zz}^\bullet = G^\bullet (I_1^\bullet - 2\lambda^{\prime 2}) + hb^\bullet - p^\bullet ,
\end{equation}
see the expression of $I_1^\bullet$ in Eq.~\eqref{InvarComp}.
The resulting normal force $\mathcal{N} = \pi \bar{\sigma}_{zz} a^2$ can be deduced from this expression by integration over the cylinder's cross section. Here, the average traction is decomposed into a weighted sum, $\bar{\sigma}_{zz} = n^\text{(f)}\bar{\sigma}_{zz}^\text{(f)} + n^\text{(s)}\bar{\sigma}_{zz}^\text{(s)}$, which can be abbreviated as $\sum_\bullet n^\bullet\bar{\sigma}_{zz}^\bullet$ for convenience, where the tractions $\bar{\sigma}_{zz}^\bullet$ are phase-related averages. In each phase, the average traction reads
\begin{equation}
    \bar{\sigma}_{zz}^\bullet = G^\bullet \lambda^2 + G^\bullet \gamma^2 \big((\alpha^\bullet)^2 + (\beta^\bullet)^2 / n^\text{(f)}\big) + hb^\bullet - p^\bullet .
    \label{EffectiveN}
\end{equation}
Overall, the effective pressure applied to the cylinder's cross section is the average $\bar p$ introduced above, and the effective magnetic induction is the average $\bar b = n^\text{(f)} b^\text{(f)} + n^\text{(s)} b^\text{(s)}$, that is, $\bar b = \bar \mu h$, according to the relationship $b^\bullet = \mu^\bullet h$ and Eq.~\eqref{Effective}.

Since both phases are isotropic (i.e., the shear moduli satisfy $\tilde{G}^\bullet = \bar{G}^\bullet$), we can replace $p^\bullet$ by $q^\bullet$ in Eq.~\eqref{EffectiveN}, see Eq.~\eqref{SubsBotton}. It follows that we can perform the substitution $\bar p = \bar q$ in \eqref{EffTractions}, where $\bar q = n^\text{(f)} q^\text{(f)} + n^\text{(s)} q^\text{(s)}$. By direct comparison with the homogenous TI case \eqref{TShear}, we have shown that the extensional and out-of-plane shearing motion of an infinite composite cylinder produces identical tractions on the cylinder's lateral boundary than the same motion applied to a specific homogeneous TI cylinder. Furthermore, using \eqref{Effective} and the following relationship (obtained using symbolic calculus software)
\begin{equation}
    \tilde{G} = n^\text{(f)} (\alpha^\text{(f)})^2 G^\text{(f)} + n^\text{(s)} \big((\alpha^\text{(s)})^2 + (\beta^\text{(s)})^2 / n^\text{(f)} \big) G^\text{(s)} ,
    \label{Effective2}
\end{equation}
it can be shown after some algebra that for a given constant vertical magnetic strength, the normal force $\mathcal N$ applied to the composite cylinder's cross section is identical to that in the homogenous TI cylinder \eqref{NShear}-\eqref{NormalForce}, with $b$ replaced by $\bar b$.

The above homogenisation result is not equivalent to that in \citet{rudykh13}, given that homogeneous deformations in each phase were assumed therein. To establish potential connections with the present study, let us compute the volume average of deformation gradient in the cylinder. First, we note that we can express the average deformation gradient $\bar{\bm F} = \sum_{\bullet} n^\bullet \bar{\bm F}^\bullet$ in the form of a weighted sum involving phase-related averages, $\bar{\bm F}^\bullet$. Special care must be taken in the computation of these averages given that the coefficients \eqref{FShearComp} are the components of a tensor with respect to a space-dependent basis (cylindrical coordinates). Hence, we need to convert this tensor representation to Cartesian coordinates prior to spatial averaging, starting from Eq.~\eqref{ShearComp} with $R^2 = X^2 + Y^2$ and $R \cos\Theta = X$, to find
\begin{equation}
    \bm{F}^\bullet = \begin{bmatrix}
        \lambda' & 0 & 0\\
        0 & \lambda' & 0\\
        F_{ZX} & F_{ZY} & \lambda
    \end{bmatrix} , \quad
    F_{Z\diamond}^\bullet = \gamma \frac{\partial}{\partial \diamond} \left(X \frac{f^\bullet(R)}{R}\right) ,
    \label{FShearCart}
\end{equation}
where $\diamond = X, Y$, in Cartesian coordinates. The final expression of the above coefficients is deduced from Eq.~\eqref{deBottonF}.

For an infinite cylinder, the volume averages of magnetic field and of magnetic induction are given by the expressions $\bar{\bf h} = h \hat{\bf e}_z$ and $\bar{\bf b} = \bar b \hat{\bf e}_z$, respectively. The average deformation gradient in each phase can be obtained by integration of \eqref{FShearCart} along a horizontal surface and over $z$, which results in a macroscopic deformation described by the phase-related average tensors
\begin{equation}
    \bar{\bm F}^\bullet = \begin{bmatrix}
        \lambda' & 0 & 0\\
        0 & \lambda' & 0\\
        \gamma \alpha^\bullet & 0 & \lambda
    \end{bmatrix} ,
\end{equation}
in Cartesian coordinates. Thus, in a cylindrical coordinate system, the deformation is described by tensors $\bar{\bm F}^\bullet$ of the form \eqref{FShear}, with the amount of shear $\gamma$ replaced by $\gamma \alpha^\bullet$. Given that $\sum_{\bullet} n^\bullet \alpha^\bullet = 1$, see Eq.~\eqref{deBotton}, we note that the average deformation gradient $\bar{\bm F}$ corresponds to that of a simple shear and extension motion with the amount of shear $\gamma$, as introduced in Eqs.~\eqref{Shear}-\eqref{FShear}. This deformation can be imposed inside a heterogeneous cylinder by enforcing a simple shear and extension motion with distinct amounts of shear in each phase.

Now, let us compute the average magneto-elastic energy $\bar \Psi = \sum_\bullet n^\bullet \bar\Psi^\bullet$ in the composite. The average energy $\bar\Psi^\bullet$ in each phase corresponds to that of an isotropic medium, given by \eqref{EnergyTI} with a single shear modulus $G^\bullet$, magnetic permeability $\mu^\bullet$, and the phase-related average invariants
\begin{equation}
    \bar I_1^\bullet = \lambda^2 + 2 \lambda^{\prime 2} + \gamma^2 \big((\alpha^\bullet)^2 + (\beta^\bullet)^2 / n^\text{(f)}\big) , \quad
    \bar I_5^\bullet = (b^\bullet)^2 ,
    \label{InvarAver}
\end{equation}
which were obtained from \eqref{InvarComp}-\eqref{deBottonF} by spatial averaging. After some manipulations, it can be shown that the resulting expression of $\bar\Psi$ matches the energy \eqref{EnergyTI} of the homogeneous TI cylinder subjected to the motion \eqref{FShear}, based on the average magnetic induction $\bar b = \bar \mu h$ and the connections \eqref{Effective}-\eqref{Effective2}, see Appendix~\ref{app:I1} for a detailed derivation.

In summary, we have shown that under an out-of-plane shear motion with extension and a vertical magnetic field, a composite cylinder made of specific magneto-elastic materials behaves identically to a homogeneous TI cylinder. The associated energy function is given by Eq.~\eqref{EnergyTI}, with the effective shear moduli $\bar G$, $\tilde G$ and magnetic permeability $\bar\mu$ defined in Eq.~\eqref{Effective}. It is worth pointing out that this result is restricted by the validity of our modelling assumptions. In fact, it does not generalise straightforwardly to other material models (see Appendix~\ref{app:I2}), magneto-deformations (Appendix~\ref{app:Torsion}), and microstructures. One exception is the case of Yeoh fibres embedded within a neo-Hookean host material (Appendix~\ref{app:I1}), for which an effective energy function \eqref{EnergyYeohTI} can be obtained in a similar fashion by assuming moderate values of the Yeoh parameter.

\section{Incremental motion of a pre-stressed magneto-elastic composite}\label{sec:SoL}

\subsection{Small-on-large theory}

For this part, we follow the work by \citet{destrade11} to derive the small-on-large theory applied to the present material model. Thus, we consider a large sample of material described by the energy function \eqref{EnergyTI}, and we introduce a small increment in the displacement field, $\tilde{\bm u}$. This increment is accompanied by a small increment in the deformation gradient, $\tilde{\bm F} = \bm{H}\bm{F}$, where $\bm{H} = \text{grad}\, \tilde{\bm u}$ is the displacement gradient tensor. Due to incompressibility, the incremental displacement field $\tilde{\bm u}$ is necessarily divergence-free, and therefore, the displacement gradient tensor $\bm H$ is trace-free.

Next, we introduce the increment $\tilde{\bf b}_0$ in the Lagrangian magnetic induction, as well as the increments in the pressure $\tilde p$, and magnetic field, $\tilde{\bf h}_0$. By linearising the products, we arrive at
\begin{equation}
    \begin{aligned}
    &\tilde{\bm\sigma}^\star = \mathbb{C} : \bm{H} + \bm{\mathcal B}\, \tilde{\bf b}^\star - \tilde p \bm{I} + p \bm{H}^\text{T} , \\
    & \tilde{\bf h}^\star = \bm{H} : \bm{\mathcal B} + \bm{M} \tilde{\bf b}^\star , 
    \end{aligned}
    \label{SoL}
\end{equation}
together with the coefficients
\begin{equation}
    \begin{aligned}
    &\mathbb{C}_{ijk\ell} = F_{j \alpha}\frac{\partial^2 \Psi}{\partial F_{i\alpha} \partial F_{k\beta}} F_{\ell\beta} = \delta_{ik} \sigma^\text{E}_{j\ell} + 4\Psi_{77} {\rm n}_i{\rm n}_j{\rm n}_k{\rm n}_\ell , \\
    &\mathcal{B}_{ijk} = F_{j \alpha}\frac{\partial^2 \Psi}{\partial F_{i\alpha} \partial \text{b}_{0\beta}} F^{-1}_{\beta k} = 2\Psi_5 (\delta_{ik}{\rm b}_j + {\rm b}_i\delta_{jk}) , \\
    &M_{ij} = F^{-1}_{\alpha i}\frac{\partial^2 \Psi}{\partial \text{b}_{0\alpha} \partial \text{b}_{0\beta}} F^{-1}_{\beta j} = 2\Psi_5\delta_{ij} ,
    \end{aligned}
    \label{SoLTens}
\end{equation}
see also \citet{karami19} for similar derivations. The starred quantities in \eqref{SoL} are pushed-forward Lagrangian increments, i.e., $\tilde{\bf h}^\star = \bm{F}^{-\text{T}\,}\tilde{\bf h}_0$ and $\tilde{\bf b}^\star = \bm{F}\tilde{\bf b}_0$. The coefficients \eqref{SoLTens} of the small-on-large linearisation are deduced from the constitutive law \eqref{EnergyTI}-\eqref{ConstitutiveExpr}. Their expression involves Kronecker's symbol $\delta_{ij}$, the coefficients $\Psi_n$ defined in Eq.~\eqref{Psi7Coeff}, and their derivatives $\Psi_{nn} = \partial \Psi_n / \partial I_n$.
These formulas are consistent with \citet{destrade11} in the isotropic limit where $\Psi_7$ and $\Psi_{77}$ vanish.

If we account for inertia, the incremental equations of motion read
\begin{equation}
    \text{div}\, \tilde{\bm\sigma}^\star = \bar\rho \ddot{\tilde{\bm u}}, \quad \text{curl}\, \tilde{\bf h}^\star = {\bf 0}, \quad \text{div}\, \tilde{\bf b}^\star = 0 ,
    \label{EqIncremental}
\end{equation}
where 
\begin{equation}
    \bar \rho = n^\text{(f)} \rho^\text{(f)} + n^\text{(s)} \rho^\text{(s)}
    \label{EffectiveRho}
\end{equation}
represents the mass density (i.e., the volume average of the mass densities $\rho^\bullet$ in each phase).
Using the homogeneity of the pre-deformation and of the magnetic field (i.e., in the case where $\bm F$, $\bf h$ and $\bf b$ are assumed constant), we deduce from the equilibrium equations that $p$ must remain constant as well. It follows that the tensor $p \bm{H}^\text{T}$ is divergence-free due to incompressibility and the equality of mixed partials, so that it can be ignored in Eq.~\eqref{SoL} under these assumptions.

\begin{remark}
    In the absence of a magnetic field (${\bf h} = {\bf 0}$), Eq.~\eqref{SoLTens} recovers known expressions of the acoustic tensor of soft fibre-reinforced composites \citep{vinh13, galich17}. Furthermore, in the undeformed state $(\bm{F} = \bm{I})$, we recover linear TI elasticity, see Eq.~(2.15) of \citet{chadwick93}.
\end{remark}

\subsection{Plane perturbations}

We seek solutions in the form of progressive plane waves defined by the unknowns
\begin{equation}
    \tilde{\bm u} = \check{\bm u}(\xi), \quad \tilde{\bf b}^\star = \check{\bf b}^\star(\xi), \quad \text{etc.}, \quad
    \xi = \bm{d}\cdot \bm{x} - c t ,
\end{equation}
which are expressed in terms of the coordinate $\xi$. As time increases, such perturbations travel with speed $c$ (to be determined), along the propagation direction indicated by the unit vector $\bm d$.

From the incompressibility condition and the incremental equations of motion \eqref{EqIncremental}, we readily deduce the orthogonality conditions $\check{\bm u}\cdot{\bm d} = 0$ and $\check{\bf b}^\star\cdot{\bm d} = 0$. Furthermore, one can derive an eigenvalue problem for the squared velocity based on the definition of a generalised \emph{acoustic tensor} $\widehat{\bm \Gamma}$, which takes the form \citep{destrade11}
\begin{equation}
    \widehat{\bm \Gamma} \check{\bm u} = \bar\rho c^2 \check{\bm u} , \quad
    \widehat{\bm \Gamma} = \widehat{\bm Q} - \widehat{\bm R} \widehat{\bm M}^{-1} \widehat{\bm R}^\text{T} ,
    \label{Acoust}
\end{equation}
with $\widehat{\bm T} = \widehat{\bm I} {\bm T} \widehat{\bm I}$ for any second-order tensor $\bm T$, and
\begin{equation}
    \widehat{\bm M}^{-1} = 2\frac{(\text{tr}\widehat{\bm M})\widehat{\bm I} - \widehat{\bm M}}{(\text{tr}\widehat{\bm M})^2 - \text{tr}\widehat{\bm M}^2} , \quad \widehat{\bm I} = {\bm I} - {\bm d}\otimes {\bm d} .
\end{equation}
The tensor $\widehat{\bm I}$ represents the projection onto the plane normal to $\bm d$, which includes the vectors $\check{\bm u}$ and $\check{\bf b}^\star$. The coefficients
\begin{equation}
    Q_{ij} = \mathbb{C}_{iIjJ}d_Id_J, \quad R_{ij} = \mathcal{B}_{iIj} d_I , \quad M_{ij} ,
    \label{AcoustEnd}
\end{equation}
are deduced from Eq.~\eqref{SoLTens}. Non-trivial eigenvector solutions $\check{\bm u} \neq {\bf 0}$ that correspond to propagating perturbations must satisfy the \emph{strong ellipticity} condition
\begin{equation}
    \bar\rho c^2 = \check{\bm u}\cdot \widehat{\bm \Gamma} \check{\bm u} > 0 , \quad \|\check{\bm u}\| = 1 ,
\end{equation}
without loss of generality. This inequality enforced for all orthogonal unit vectors $\bm d$ and $\check{\bm u}$ coincides with the \emph{strict hyperbolicity} condition. Under this condition, transverse waves can propagate in any direction, and their speed $c$ is a real number.

In the present case, using \eqref{Acoust}-\eqref{AcoustEnd} and the property $\widehat{\bm I}{\bm d} = \bm{0}$, we find that $\widehat{\bm M} = 2 \Psi_5 \widehat{\bm I}$, and that $\widehat{\bm R} = ({\bm d}\cdot {\bf b}) \widehat{\bm M}$ is a symmetric tensor. Thus, we arrive at the expression of the generalised acoustic tensor
\begin{equation}
    \begin{aligned}
    &\widehat{\bm \Gamma} = a_1 \widehat{\bm I} + a_2\, (\widehat{\bm I} {\bf n}) \otimes (\widehat{\bm I} {\bf n}) ,\\
    &a_1 = {\bm d}\cdot ({\bm \sigma}^\text{E} - {\bf h}\otimes {\bf b}){\bm d} , \quad a_2 = 4 \Psi_{77} ({\bm d}\cdot{\bf n})^2 ,
    \end{aligned}
    \label{WaveCoeff}
\end{equation}
where ${\bf h} = {\bf b}/\bar\mu$ and $\Psi_{77} = \frac34 (\bar G - \tilde G) I_7^{-5/2}$.
This formula is identical to that obtained in the magnetically inert case by \citet{galich17}. Solving for the roots of the acoustic tensor's characteristic polynomial, we observe that $\widehat{\bm \Gamma}$ has at most two non-trivial eigenvalues
\begin{equation}
    \bar\rho c^2 = a_1 , \quad \bar\rho c^2 = a_1 + a_2\, ({\bf n}\cdot  \widehat{\bm I}{\bf n}) ,
    \label{WaveSpeed}
\end{equation}
from which the wave speed $c$ is deduced.

For waves propagating along the fibre direction, the projection $\widehat{\bm I}{\bf n}$ of the fibre direction onto the polarisation plane vanishes, and both eigenvalues are equal to $a_1$, corresponding to shear waves polarised transversely to the fibres. The scalar product $\bm{d}\cdot{\bf n}$ vanishes for waves propagating transversely to the fibre direction and with either a vertical or a horizontal polarisation, so that both eigenvalues are equal to $a_1$ as well. Otherwise, these eigenvalues are distinct, and they correspond to shear waves polarised transversely or along the direction indicated by $\widehat{\bm I}{\bf n}$, respectively.

The incremental magnetic induction $\check{\bf b}^\star$ is then proportional to $({\bf b}\cdot{\bm d}) \partial_\xi \check{\bm u}$, see the derivation in \citet{destrade11}. Therefore, the vector $\check{\bf b}^\star$ is aligned with the wave polarisation, and there is no incremental magnetic induction if the waves propagate in a direction normal to ${\bf b}$.

\subsection{The angled shear wave identity}

\citet{li20} proposed a formula that connects the state of static stress in a material to the speed of two shear waves propagating in orthogonal directions. Based on the small-on-large theory, this formula is potentially useful in material testing and inspection. Here, we propose to extend this result to fibre-reinforced magneto-elastic solids described by the constitutive theory \eqref{EnergyTI}.

To do so, we assume that the deformation gradient tensor is represented by a diagonal matrix, $\bm{F} = \text{diag}[\lambda_x, \lambda_y, \lambda_z]$, where the constant coefficients $\lambda_i$ are the stretch ratios. This motion corresponds to the deformation of a cube into a cuboid, with stretch ratio $\lambda_x$ along the $x$-axis, etc. Thus, we have ${\bf n} = \lambda_z\hat{\bf n}_0$, where ${\bf n}$ indicates the current orientation of the fibres, which is aligned with the vertical axis, $\hat{\bf n}_0 = \hat{\bf e}_z$. From $I_7 = \lambda_z^2$, we deduce the expression of the coefficients $\Psi_7$, $\Psi_{77}$ arising in the above acousto-elasticity formulas \eqref{WaveCoeff}-\eqref{WaveSpeed}, see also Eq.~\eqref{Psi7Coeff}. Here, the material is subjected to a uniform vertical magnetic field, ${\bf h} = h\hat{\bf e}_z$, and the pressure $p$ is assumed constant.

Now, let us substitute the extra stress $\bm{\sigma}^\text{E}$ for $\bm{\sigma} + p \bm{I}$ in Eq.~\eqref{WaveCoeff}. The constitutive law \eqref{ConstitutiveExpr} tells us that the total stress is a diagonal tensor, $\bm{\sigma} = \text{diag}[\sigma_x, \sigma_y, \sigma_z]$, whose principal stresses $\sigma_i$ are constants. Thus,
\begin{equation}
    a_1 = p + \sigma_id_i^2 - hb d_z^2 , \quad a_2 = 4\Psi_{77} \lambda_z^2 d_z^2 ,
    \label{AngleCoeff}
\end{equation}
where we have used the representation ${\bm d} = d_i \hat{\bf e}_i$ for the unit vector defining the direction of propagation, with summation over repeated subscripts. Here, the projection $\widehat{\bm I}{\bf n}$ is proportional to the vector $\hat{\bf e}_z - d_z\bm{d}$, and the coefficient ${\bf n}\cdot \widehat{\bm I}{\bf n}$ in Eq.~\eqref{WaveSpeed} equals $\lambda_z^2 (1 - d_z^2)$. For later use, we replace the notation $c$ for the wave speed by $c_{\check{\bm u}|{\bm d}}$, whose expression can be deduced from the above formulas.

To derive the \emph{angled shear wave identity}, we select the unit vectors $\bm d$ and $\check{\bm u}$ within a plane formed by two principal axes of deformation, e.g., the horizontal plane $z=0$ or the vertical plane $y=0$ (the vertical plane $x=0$ yields similar properties). Then, we swap the propagation direction $\bm d$ for the polarisation direction $\check{\bm u} = \check{u}_i \hat{\bf e}_i$ and compute the difference between the squared wave speeds $c_{\check{\bm u}|{\bm d}}$ and $c_{{\bm d}|\check{\bm u}}$ obtained for each wave. For many materials, this difference is directly related to the principal stresses, see for instance \citet{berjamin24}. When swapping the propagation and polarisation directions, we note that it suffices to substitute $\bm d$ with $\check{\bm u}$ in Eqs.~\eqref{WaveCoeff}-\eqref{AngleCoeff}.

\begin{figure}
    \centering
    \includegraphics{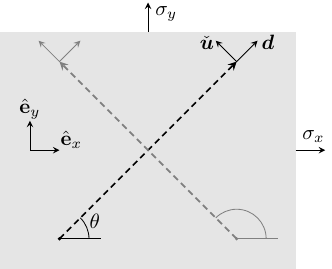}
    \caption{Angled shear wave identity. A rectangular cuboid is maintained in a state of static stress (sectional view for constant $z$). Shear waves propagate along two directions with in-plane polarisation.}
    \label{fig:angle}
\end{figure}

Let us now consider each case separately.
\begin{itemize}
\item If $\bm d$, $\check{\bm u}$ belong to the horizontal plane $z=0$, then the scalar products $\bm{d}\cdot{\bf n}$ and $\check{\bm u}\cdot{\bf n}$ vanish. Therefore, using the above formulas, we note that the squared wave speeds satisfy
\begin{equation}
    \bar\rho c_{\check{\bm u}|{\bm d}}^2 - \bar\rho c_{{\bm d}|\check{\bm u}}^2 = (\sigma_x - \sigma_y)\cos(2\theta) ,
    \label{AngledLi}
\end{equation}
where we have set $d_x = \cos \theta$ and $\check{u}_x = -\!\sin \theta$. Thus, we recover the classical identity derived by \citet{li20}. Eq.~\eqref{AngledLi} is a \emph{universal relation}, given that it does not explicitly depend on the material's constitutive law (it is valid for a whole class of materials). An illustration for these wave speed measurements is given in Figure~\ref{fig:angle}.
\item If $\bm d$, $\check{\bm u}$ belong to the vertical plane $y=0$, then
\begin{equation}
    \bar\rho c_{\check{\bm u}|{\bm d}}^2 - \bar\rho c_{{\bm d}|\check{\bm u}}^2 = (\sigma_x - \sigma_z + hb)\cos(2\theta) ,
    \label{AngledMod}
\end{equation}
where we have used similar notations. Therefore, we obtain a modified identity whenever the magnetic field is not orthogonal to the polarisation and propagation vectors. Note in passing that the relationship \eqref{AngledMod} is no longer universal, since it involves the magneto-elastic response via $b = \bar \mu h$.
\end{itemize}
Finally, the present magneto-elastic material behaviour entails a modification of the classical identity \eqref{AngledLi}. Nevertheless, if the magnetic field is known, then we can still obtain an estimation of the difference in the total stress along two orthogonal directions based on the above formulas.

\subsection{Uniaxial tension-compression}\label{subsec:Tensile}

We consider a static uniaxial motion along the vertical axis of a composite cylinder, which is aligned with the initial direction of the fibre. This pre-deformation is described by the diagonal deformation gradient tensor \eqref{FShear} with $\gamma = 0$, where $\lambda_z = \lambda$ is the vertical stretch and $\lambda_x = \lambda_y$ is the radial stretch, $\lambda' = \lambda^{-1/2}$.

As done earlier, we assume that the material is subjected to a uniform vertical magnetic field. Therefore, the vertical normal force in this material is given by Eqs.~\eqref{NShear}-\eqref{NormalForce}, with the amount of shear $\gamma = 0$. If we consider a single cylinder subjected to an external radial load $\sigma_n = \sigma_{rr}|_{r=a}$ on the lateral boundary ($a = A\lambda'$), then we deduce the value of $q$ from the expression of the radial stress \eqref{TShear}, which entails the relationship
\begin{equation}
    \begin{aligned}
    &\mathcal{F} = \lambda - \lambda^{-2} + \mathcal{F}_h / \lambda , \\
    &\mathcal{F} = \mathcal{N}/(\pi A^2\bar G), \quad \mathcal{F}_h = (\bar\mu h^2 + \sigma_n) / \bar G ,
    \end{aligned}
    \label{InterStretch}
\end{equation}
where $\mathcal{F}$ and $\mathcal{F}_h$ represent the normalised vertical stress and its magnetic part, respectively. For instance, if the cylinder is in vacuum, then the boundary conditions require that $\sigma_n$ equals $-\mu_0 h^2/2$, see Eq.~\eqref{StressMax}. For a given value of the total vertical force $\mathcal{F}$ and of the magnetic force $\mathcal{F}_h$, the vertical stretch $\lambda > 0$ can be deduced from the relationship \eqref{InterStretch}, which takes the form of a cubic polynomial equation in $\lambda$. In the special case of a magnetically inert cylinder in vacuum ($\mathcal{F}_h = 0$), the stretch ratio is only function of the vertical force $\mathcal{F}$.

\begin{figure}
    \centering
    \includegraphics{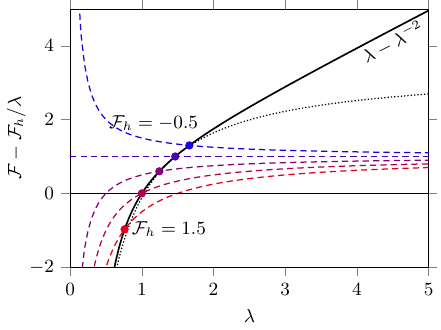}
    \caption{Operating point \eqref{InterStretch} of a magneto-elastic composite cylinder under uniaxial loading for a controlled normal force such that $\mathcal{F} = 1$. The magnetic part of the normal force takes the values $\mathcal{F}_h = \lbrace -0.5, 0, \dots, 1.5 \rbrace$. The dotted line marks the approximation \eqref{Taylor}. \label{fig:StretchMag}}
\end{figure}

We illustrate the determination of the vertical stretch ratio based on Eq.~\eqref{InterStretch} in Figure~\ref{fig:StretchMag}, where we have set $\mathcal{F} = 1$ and where $\mathcal{F}_h$ is varied from $-0.5$ to $1.5$ by steps of $0.5$. Thus, this example corresponds to a configuration where the vertical normal force $\mathcal{F}$ is controlled. Using symbolic calculus software, we find the approximate solution
\begin{equation}
    \lambda \simeq 1.46557 - 0.417238 \, \mathcal{F}_h + 0.611492 \left(\mathcal{F}-1\right) ,
    \label{Taylor}
\end{equation}
based on a truncated Taylor series expansion of the roots of \eqref{InterStretch}\textsubscript{1} about $(\mathcal{F}_h, \mathcal{F}) \simeq (0,1)$. This approximation is represented using a dotted line in Fig.~\ref{fig:StretchMag}, where its good accuracy for $\mathcal{F}=1$ and the selected values of $\mathcal{F}_h$ is demonstrated. We note that an increase in $\mathcal{F}_h$ entails a decrease in the stretch $\lambda$, whereas variations of the total force $\mathcal{F}$ have the opposite effect.

A relationship similar to \eqref{InterStretch} was obtained by \citet{padmanabhan24} in the case of traction-free boundary conditions defined by $\sigma_n = 0$ and $\mathcal{N}=0$, where the Gent model of hyperelasticity was combined to a hard-magnetic material model. In this case ($\mathcal{F}=0$), the relationship \eqref{InterStretch} is a depressed cubic equation that can be solved directly using Cardano's formula
\begin{equation}
    \lambda = \sqrt[3]{\tfrac12 + \sqrt{\tfrac14 + \tfrac1{27}\mathcal{F}_h^3}} + \sqrt[3]{\tfrac12 - \sqrt{\tfrac14 + \tfrac1{27}\mathcal{F}_h^3}} ,
\end{equation}
for $\mathcal{F}_h > \sqrt[3]{-27/4}$, that is $\mathcal{F}_h > -1.89$, approximately. Therefore, the material undergoes a contraction $\lambda<1$ along the vertical axis if $\mathcal{F}_h > 0$, and it undergoes a vertical elongation otherwise. Note in passing that negative values of $\mathcal{F}_h$ are hardly achievable in soft magneto-active materials according to \eqref{InterStretch}, unless a large compressive lateral force is applied ($\sigma_n < 0$). Here too, an increase in $\mathcal{F}_h$ entails a decrease in the stretch $\lambda$.

In the case of a rectangular slab of composite material described by the TI energy function \eqref{EnergyTI} and subjected to the same static loading, we find that the Lagrange multiplier $p$ must be constant as well. The non-zero tractions on the lateral boundaries $\sigma_{x} = \sigma_{y}$ have the same form as the radial traction $\sigma_{rr}$ in Eq.~\eqref{TShear}, while the normal traction $\sigma_{z}$ is given by the same formula as Eq.~\eqref{NShear} with $\gamma=0$. The vertical normal force is given by $ \mathcal{N} = \sigma_{z} a_x a_y$ instead of \eqref{NormalForce}, where $a_x = A_x\lambda'$ and $a_y = A_y\lambda'$ denote the dimensions of the deformed slab along the $x$ and $y$ directions.

If we consider that the slab is subjected to an external normal load $\sigma_n$ on the lateral boundaries, then we deduce the value of $q$ from the expression of the tractions $\sigma_{x}$ and $\sigma_{y}$, which entails the relationship \eqref{InterStretch}, with \eqref{InterStretch}\textsubscript{2} replaced by $\mathcal{F} = \mathcal{N}/(A_xA_y\bar G)$. If the slab is in vacuum \eqref{StressMax}, then we recover the expression $\sigma_n = -\mu_0 h^2/2$ of the lateral tractions. To conclude, the vertical stretch ratio in a slab under uniaxial loading can still be deduced from the knowledge of the vertical normal force and of its magnetic part, as illustrated in Fig.~\ref{fig:StretchMag}. In what follows, we investigate shear wave propagation properties and the macroscopic stability of such a deformed structure.

\paragraph{Waves.}

For a TI material described by the energy function \eqref{EnergyTI} and subjected to an uniaxial pre-deformation of stretch $\lambda = (\lambda')^{-2}$ along the fibers (aligned vertically), we deduce the expression of the wave speeds $c$ from the formulas \eqref{WaveSpeed}-\eqref{AngleCoeff} with the principal stresses \eqref{TShear}-\eqref{NShear}, the invariant $I_7 = \lambda^2$, the scalar product ${\bf n}\cdot \widehat{\bm I}{\bf n} = \lambda^2 (1 - d_z^2)$, and the relationship \eqref{SubsBotton}:
\begin{equation}
    \begin{aligned}
        \bar\rho c^2 &= \bar{G} \lambda^2 d_z^2 + (\tilde G - \bar{G} d_z^2)/\lambda , \\
        \bar\rho c^2 &= \bar{G} \lambda^2 d_z^2 + (\tilde G - \bar{G} d_z^2)/\lambda + 2E d_z^2 (1 - d_z^2) /\lambda ,
    \end{aligned}
    \label{TensWaves}
\end{equation}
where $E = \frac32 (\bar G - \tilde G) \geq 0$ is the degree of anisotropy. As shown in \citet{galich17}, we observe that these two wave speeds are equal if ${\bm d}$ is orthogonal to or aligned with the fibres ($d_z^2 = 0, 1$). Otherwise, the wave speed deduced from the second expression in \eqref{TensWaves} is always greater or equal to that deduced from the first one, leading to a dependence of the wave speed with the direction of propagation when the material is deformed (or when it is subjected to a permanent vertical magnetic field). In the undeformed material ($\lambda = 1$), waves propagating along the fibres or transversely to them travel at the same constant speed, $\sqrt{\tilde{G}/\bar\rho}$.

If we consider a controlled experiment in which the vertical normal force is such that $\mathcal{F} = 1$ and the lateral tractions $\sigma_n$ are known, then we deduce from Eq.~\eqref{InterStretch} that the stretch is function of $h$ and of the material parameters $\bar G$, $\tilde G$, $\bar \mu$ introduced in Eq.~\eqref{Effective}. Using the Taylor series approximation \eqref{Taylor}, the wave speeds \eqref{TensWaves} can be approximated using
\begin{equation}
    \lambda^2 \simeq 2.15 - 1.22\, \mathcal{F}_h , \quad \lambda^{-1} \simeq 0.682 + 0.194\, \mathcal{F}_h ,
    \label{TensWavesTaylor}
\end{equation}
where $\mathcal{F}_h$ has a sufficiently small absolute value (the validity of this approximation for $-0.5 \leq \mathcal{F}_h \leq 1.5$ is illustrated in Fig.~\ref{fig:StretchMag}). Given the definition of $\mathcal{F}_h$ in Eq.~\eqref{InterStretch}, we have therefore expressed the wave speeds in terms of the magnetic field and of the material parameters.

\begin{figure}
    \centering
    \includegraphics{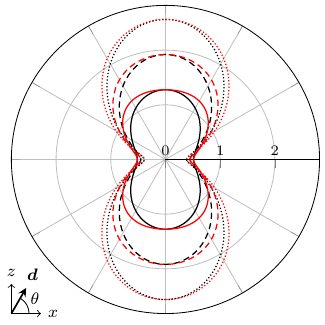}
    \caption{Dependence of the normalised wave speeds $\bar\rho c^2/\bar G$ from \eqref{TensWaves}-\eqref{TensWavesTaylor} with respect to the vertical component of the propagation vector (black: \eqref{TensWaves}\textsubscript{1}, red: \eqref{TensWaves}\textsubscript{2}). The vertical normal force satisfies $\mathcal{F}=1$ while its magnetic part takes the values $\mathcal{F}_h = \lbrace -0.5, 0, 0.5 \rbrace$, corresponding to dotted, dashed and solid lines, respectively. \label{fig:Polar}}
\end{figure}

We illustrate the dependence of the wave speed with respect to the propagation direction in Figure~\ref{fig:Polar}. In this polar plot, we represent the evolution of the normalised wave speeds $\bar\rho c^2/\bar G$ deduced from Eqs.~\eqref{TensWaves}-\eqref{TensWavesTaylor} for $\tilde G / \bar G = 0.66$ with respect to the angle parameter $\theta$ such that $d_z = \sin\theta$. The black curves correspond to the first expression of $\bar\rho c^2$ in Eq.~\eqref{TensWaves} (i.e., shear waves polarised transversely to the projection of $\hat{\bf e}_z$ onto the plane orthogonal to $\bm d$), whereas the lighter red curves correspond to the second expression (shear waves polarised along this projection). Coherently with earlier observations, we note that the latter wave velocity is always greater or equal to the former one. By varying the magnetic part $\mathcal{F}_h$ of the vertical normal force, we note that the magnitude of these shear wave speeds can be modified. In addition, the difference between these two wave speeds varies with the applied magnetic force as well.

\paragraph{Stability.}

The expression of the shear wave speeds in \eqref{TensWaves} yields an explicit condition for the macroscopic stability of the material in uniaxial contraction \citep{galich17}. In fact, setting $d_z = 1$, we note that the wave speed is real provided that $\lambda$ is greater than a critical stretch $\lambda_\text{c} \leq 1$, which is given by the expression
\begin{equation}
    \lambda_\text{c} = (1-\tilde{G}/\bar{G})^{1/3} , \quad 0 \leq \tilde{G}/\bar{G} \leq 1 .
    \label{Stability}
\end{equation}
Therefore, the range of stretches covered in Fig.~\ref{fig:StretchMag} should potentially be restricted. To illustrate this process, if $\tilde{G}/\bar{G} = 0.66$, then the critical stretch reads $\lambda_\text{c} \simeq 0.7$. Stretches smaller than this value will induce macroscopic instability.

Now, let us consider again a controlled experiment in which the vertical normal force is such that $\mathcal{F} = 1$ and the lateral tractions $\sigma_n$ are known. According to Eq.~\eqref{InterStretch}, the magnetic force $\mathcal{F}_h = \lambda - \lambda^2 + \lambda^{-1}$ is a decreasing function of the vertical stretch $\lambda > 0$. Therefore, the stability requirement $\lambda > \lambda_\text{c}$ can be rewritten as
\begin{equation}
    \mathcal{F}_h < \lambda_\text{c} - \lambda_\text{c}^2 + \lambda_\text{c}^{-1} ,
    \label{Stability2}
\end{equation}
where the right-hand side is greater or equal to unity. Using \eqref{Stability} and the definition of $\mathcal{F}_h$ in Eq.~\eqref{InterStretch}, this restriction gives us an upper bound for the magnitude $|h|$ of the magnetic field in terms of the material parameters $\bar G$, $\tilde G$, $\bar \mu$ introduced in Eq.~\eqref{Effective}. If $\tilde{G}/\bar{G} = 0.66$, then we obtain the bound $\mathcal{F}_h < 1.64$. Hence, for this choice of parameters, the range of stretches covered in Fig.~\ref{fig:StretchMag} does not induce macroscopic instability. For a study of material instability under other loading scenarios, the interested reader is referred to \citet{goshkoderia17}.

\section{Antiplane shear waves in pre-stressed fibre composites}\label{sec:2D}

In the present section, the material is subjected to a large uniaxial deformation of axial stretch $\lambda = (\lambda')^{-2}$, which is induced by a uniform vertical magnetic field ${\bf h} = h\hat{\bf e}_z$. Here, we consider time-harmonic antiplane shear waves polarised along the vertical axis, i.e., the incremental displacement field in Eqs.~\eqref{SoL}-\eqref{EqIncremental} is such that $\tilde{\bm u} = \tilde{u}(x,y) \text{e}^{-\text{i}\omega t} \hat{\bf e}_z$, and $\omega$ denotes the angular frequency. The incremental quantities $\tilde p$, $\tilde{\bf h}^\star$, $\tilde{\bf b}^\star$ do not depend on $z$. Thus, using the incompressibility property and the equality of mixed partials, the incremental equations of motion become
\begin{equation}
    \nabla\cdot (\tilde G \lambda^{\prime 2}  \nabla \tilde{u}) + \bar\rho \omega^2 \tilde{u} = 0, \quad \nabla\cdot (\bar\mu \nabla \tilde{\phi}) = 0 ,
    \label{SoL2D}
\end{equation}
where $\nabla$ is the gradient operator. The relationship $\tilde{\bf h}^\star = h\nabla \tilde{u} - \nabla\tilde{\phi}$ defines a scalar magnetic potential $\tilde\phi$, which satisfies the identity $\tilde{\bf b}^\star = -\bar\mu \nabla \tilde\phi$.

\begin{figure}
    \centering
    \includegraphics{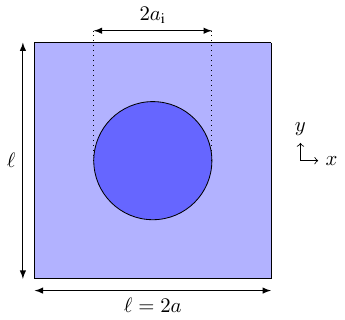}
    \caption{Periodic unit cell of a composite material (cross-section), with internal radius $a_\text{i}$ and cell size $2a$ after homogeneous tensile deformation along the fibre direction.
    \label{fig:Periodic}}
\end{figure}

We consider a periodic composite material made of a cylindrical fibre embedded inside a matrix. In fact, we will consider a perfectly bonded composite made of two isotropic neo-Hookean elastomers with shear modulus $\tilde G = \bar G$. The fibre located in the region $0\leq R\leq A_\text{i}$ is a magneto-elastic material described by the energy function \eqref{EnergyTI} with $\tilde G = G^\text{(f)}$, $\bar\mu = \mu^\text{(f)}$ and mass density $\rho^\text{(f)}$, whereas the matrix located in the region $A_\text{i} \leq R$ such that $|X|, |Y| \leq A$ has parameters $\tilde G = G^\text{(s)}$, $\bar\mu = \mu^\text{(s)}$ and mass density $\rho^\text{(s)}$, see Figure~\ref{fig:Periodic}. The deformed inner and outer dimensions of the unit cell equal $a_\text{i} = A_\text{i}\lambda'$ and $a = A\lambda'$, respectively. Suitable boundary conditions are enforced at the inner and outer interfaces, allowing us to determine the jump $p^\text{(f)}-p^\text{(s)}$ in the static pressure across the inner boundary. Formally, the motion is governed by Eq.~\eqref{SoL2D} in each phase, with the required boundary conditions at the inner and outer interfaces of the unit cell.

\subsection{Results from periodic homogenisation}

Assuming separation of scales between the macroscopic wavelength $L$ and the spatial period $\ell = 2a$, we perform an asymptotic analysis based on the small parameter $\epsilon = \ell/L$. To do so, we assume that the material parameters $G^\bullet$, $\rho^\bullet$, $\mu^\bullet$ with $\bullet \in \lbrace (\text{f}),(\text{s})\rbrace$ do not depend on $\epsilon$. Next, a `slow' and a `fast' set of spatial coordinates are introduced, namely $\bm{x}$ and $\bm{x}/\epsilon$, which are assumed independent. Solutions are sought in the form of a power series of the parameter $\epsilon \ll 1$ whose coefficients are functions of these spatial coordinates, and $\ell/\epsilon$-periodicity with respect to the fast coordinates is assumed. By matching identical orders of the small parameter $\epsilon$, a hierarchy of boundary-value problems can be derived. Based on an averaging procedure over the `fast' spatial coordinates, the homogenised equations of motion at every order can be obtained \citep{andrianov21}.

Following \citet{wautier15}, we seek $\tilde{u}$, $\tilde{\phi}$ in the form of a power series in $\epsilon$, whose coefficients satisfy a boundary-value problem defined over the unit cell{\,---\,}these coefficients are called `cell functions'. The asymptotic expansion must be completed by the governing equations for the mean fields $U = \tilde{u}|_{\epsilon \to 0}$ and $\Phi = \tilde{\phi}|_{\epsilon \to 0}$, which take the form of a time-harmonic wave equation for $U$ and of a Laplace equation for $\Phi$, owing to the invariance of the microstructure under a rotation of angle $\pi/2$ (i.e., $C_4$-symmetry). The mean field equation for $U$ describes the propagation of antiplane shear waves in the periodic composite material (with an effective wave speed), whereas the other mean field equation governs the magnetic potential $\Phi$ of these incremental waves (with an effective magnetic permeability). The effective parameters in these equations are deduced from the cell functions via spatial averaging.

Apart from Obnosov's concentric squares problem \citep{bellis20}, an exact determination of these effective coefficients is not always feasible, in general. Therefore, computational techniques might be implemented in practice, for instance based on fast Fourier transform algorithms \citep{cornaggia20, chen21}. Nevertheless, closed-form expressions for the effective coefficients can also be obtained in an approximate fashion, based on suitable assumptions. In the dilute limit, by replacing the initial unit cell with concentric cylinders of circular cross-section, application of a result by \citet{andrianov99} leads to the mean field equations
\begin{equation}
    \tilde G \lambda^{\prime 2} \nabla^2 U + \bar\rho \omega^2 U = 0, \quad \tilde\mu \nabla^2 \Phi = 0 ,
    \label{EffectiveEq}
\end{equation}
where the coefficient $\tilde G$ takes the so-called Maxwell--Garnett form \eqref{Effective}-\eqref{deBotton} and $\bar \rho$ is the average mass density \eqref{EffectiveRho}. By analogy, a similar approximate expression for the effective magnetic permeability $\tilde\mu$ would be obtained, namely
\begin{equation}
    \tilde\mu \simeq \mu^\text{(s)} \frac{(1+n^\text{(f)}) \mu^\text{(f)} + n^\text{(s)}\mu^\text{(s)}}{n^\text{(s)}\mu^\text{(f)} + (1+n^\text{(f)})\mu^\text{(s)}} = \mu^\text{(s)} \frac{1 - \Lambda\, n^\text{(f)}}{1 + \Lambda\, n^\text{(f)}} ,
    \label{EffectiveMu}
\end{equation}
where $n^\text{(f)} + n^\text{(s)} = 1$ and $\Lambda = \frac{\mu^\text{(s)}-\mu^\text{(f)}}{\mu^\text{(s)}+\mu^\text{(f)}}$. These results are consistent with \citet{joyce17} where the interested reader will find an alternative derivation of the effective stiffness based on integral equations, as well as more general estimates valid in the non-dilute case.

The effective wave equation \eqref{EffectiveEq}\textsubscript{1} is associated to the wave speed $c = \lambda' \sqrt{\tilde G / \bar\rho}$. Thus, wave dispersion is governed by the dispersion relationship $\kappa = \omega / c$, where $\kappa$ is the wave number and $\omega$ is the angular frequency. As far as Eq.~\eqref{EffectiveEq}\textsubscript{2} is concerned, the only radial solutions for $\Phi$ on an infinite plane that are not singular at the origin are constant functions. Therefore, we conclude that there is no incremental magnetic field associated with the present motion.

While the effective stiffness $\tilde G \lambda^{\prime 2}$ in \eqref{EffectiveEq} would be coherent with the TI magneto-elastic theory \eqref{EnergyTI}, it appears that our initial estimation of the effective magnetic response contradicts the present periodic homogenisation result, since $\bar\mu \neq \tilde\mu$ in general. A likely reason for this mismatch is that the out-of-plane shearing motion studied in Section~\ref{sec:Shear} involves a uniform and permanent vertical magnetic field. Therefore, this single-fibre problem does not account for potential in-plane variations of the magnetic field.

To accommodate for this mismatch, one might propose an anisotropic magnetic response based on further scalar invariants \eqref{Invar} in the expression of the TI magneto-elastic energy \eqref{EnergyTI}, such as the quantities \citep{bustamante10}
\begin{equation}
    \begin{aligned}
    &I_2 = \tfrac12 \big( I_1^2 -  \text{tr}(\bm{C}^2) \big) , \quad I_4 = {\bf b}_0 \cdot {\bf b}_0, \quad I_6 = {\bf b}_0 \cdot \bm{C}^2{\bf b}_0, \\
    &I_8 = \hat{\bf n}_0\cdot \bm{C}^2\hat{\bf n}_0 ,\quad I_{9} = \hat{\bf n}_0 \cdot {\bf b}_0 , \quad I_{10} = \hat{\bf n}_0 \cdot \bm{C}{\bf b}_0 .
    \end{aligned} \raisetag{1.2\baselineskip}
    \label{InvarExtra}
\end{equation}
For instance, the effective magneto-elastic energy of laminates includes an additional term, which involves the ratio between $I_9^2$ and the purely elastic invariant \citep{karami19}
\begin{equation}
    \hat{\bf n}_0 \cdot \bm{C}^{-1}\hat{\bf n}_0 = I_2 - I_1 I_7 + I_8 ,
    \label{InvarKarami}
\end{equation}
where use was made of the Cayley--Hamilton theorem, but this specific form of the magnetic response might not apply to fibre-reinforced composites. A suitable magneto-elastic energy would lead to an effective magnetic permeability equal to $\tilde\mu$ for in-plane magnetic fields, and to an effective magnetic permeability equal to $\bar\mu$ for out-of-plane magnetic fields oriented along the fibres. We leave this potential generalisation to the reader's curiosity.

\subsection{Results from the Bloch-Floquet analysis}

We will verify the above observations by means of the Bloch-Floquet technique, aka. the plane wave expansion method. To do so, we place ourselves in the same situation as for the periodic homogenisation analysis.

The deformed composite medium is periodic along the $x$ and $y$ axes, with period $\ell = 2a$ in both directions. Following Floquet's theorem, we seek solutions to Eq.~\eqref{SoL2D} as a product between a spatially periodic function (represented by its Fourier series) and a complex exponential function (with Floquet exponent $\bm \kappa$). Thus,
\begin{equation}
    \tilde u = \sum_{\bm G} U[\bm{G}]\, \text{e}^{\text i (\bm{G} + \bm{\kappa})\cdot \bm{x}}, \quad \tilde \phi = \sum_{\bm G} \Phi[\bm{G}]\, \text{e}^{\text i (\bm{G} + \bm{\kappa})\cdot \bm{x}},
    \label{BlochRep}
\end{equation}
where the Bloch wave vector ${\bm \kappa}$ and reciprocal lattice vectors $\bm G$ are orthogonal to the vertical $z$-axis. Reciprocal lattice vectors span the linear combinations $\bm{G} = \frac{2\pi}{\ell} m_k \hat{\bf e}_k$ with integer coefficients $(m_x,m_y)$, where summation over repeated indices is assumed.

In Eq.~\eqref{SoL2D}, we replace the strain-dependent shear modulus $\tilde G \lambda^{\prime 2}$, the mass density $\bar\rho$ and the magnetic permeability $\bar \mu$ by their Fourier series expansion,
\begin{equation}
    \tilde G \lambda^{\prime 2} = \sum_{\bm G} \mathscr{G}[\bm{G}]\, \text{e}^{\text i \bm{G}\cdot \bm{x}}, \quad \text{etc.},
\end{equation}
with the Fourier coefficients $\mathscr{G}$, $\mathscr{R}$ and $\mathscr{M}$, respectively. Then, upon projection onto the Fourier basis, we find
\begin{equation}
    \begin{aligned}
    &\sum_{\bm G} \left({\text K}_{\bm{G}'\bm{G}}  - \omega^2 {\text M}_{\bm{G}'\bm{G}}\right) U[\bm{G}] = 0, \\
    &\sum_{\bm G} {\text D}_{\bm{G}'\bm{G}}\, \Phi[\bm{G}] = 0  ,
    \end{aligned}
    \label{Bloch}
\end{equation}
where
\begin{equation}
    \begin{aligned}
    &{\text K}_{\bm{G}'\bm{G}} = (\bm{G}' + \bm{\kappa})^\text{T} \mathscr{G}[\bm{G}'-\bm{G}]\, (\bm{G} + \bm{\kappa}) ,\\
    &{\text M}_{\bm{G}'\bm{G}} = \mathscr{R}[\bm{G}'-\bm{G}] ,\\
    &{\text D}_{\bm{G}'\bm{G}} = (\bm{G}' + \bm{\kappa})^\text{T} \mathscr{M}[\bm{G}'-\bm{G}]\, (\bm{G} + \bm{\kappa}),
    \end{aligned}
\end{equation}
with the identities \eqref{Bloch} to be satisfied for all the reciprocal lattice vectors $\bm{G}'$, see for instance \citet{berjamin22} for details about the implementation.

\begin{figure}
    \centering
    \includegraphics{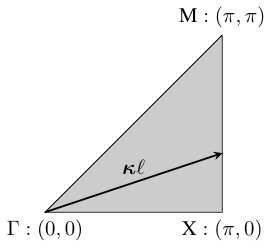}
    \vspace{1em}
    
    \includegraphics{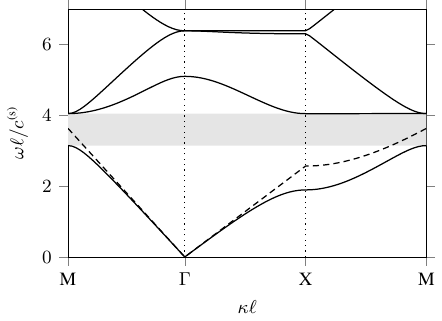}
    \caption{Soft and light fibres embedded periodically in a rubber-like host material (see Fig.~\ref{fig:Periodic}). Top: irreducible Brillouin zone. Bottom: dispersion diagram. Solid lines correspond to the wave modes obtained by using the Bloch-Floquet technique, whereas dashed lines mark the dispersion relationship deduced from the homogenisation theory.}
    \label{fig:BG1}
\end{figure}

On the one hand, Eq.~\eqref{Bloch}\textsubscript{1} can be viewed as a generalised eigenvalue problem for the squared angular frequency $\omega^2$. Indeed, the resolution of this algebraic problem provides the dispersion relationship, i.e., the relationship between the angular frequency $\omega$ and the wave vector $\bm \kappa$. This analysis can be reduced to the normalised wave vectors ${\bm\kappa} \ell$ that belong to the irreducible Brillouin zone represented in Figure~\ref{fig:BG1}-top. For each of these wave vectors, we construct the matrices $[{\text K}_{\bm{G}'\bm{G}}]$ and $[{\text M}_{\bm{G}'\bm{G}}]$ using a Fast Fourier Transform (FFT) algorithm, and we solve the eigenvalue problem \eqref{Bloch}\textsubscript{1} to find the normalised angular frequency, $\omega \ell / c^\text{(s)}$, where $c^\text{(s)}$ is the shear wave speed in the deformed host material.

On the other hand, Eq.~\eqref{Bloch}\textsubscript{2} is a system of linear algebraic equations. Non-trivial solutions can be obtained provided that the matrix $[{\text D}_{\bm{G}'\bm{G}}]$ is singular. Therefore, it suffices to construct this matrix using FFT based on a similar procedure as for the matrix $[{\text K}_{\bm{G}'\bm{G}}]$, and by selecting the same dimensionless wave vectors ${\bm\kappa} \ell$ that belong to the edges of the irreducible Brillouin zone. Then, the determination of the nullspace of $[{\text D}_{\bm{G}'\bm{G}}]$ yields the incremental magnetic potential, see Eq.~\eqref{BlochRep}.

We illustrate this procedure in Figure~\ref{fig:BG1}-bottom. Here, the host material has density $\rho^\text{(s)} = 10^3$~kg/m\textsuperscript{3} and effective shear modulus $G^\text{(s)} \lambda^{\prime 2} = 1$~kPa after deformation. The deformed fibre of radius $a_\text{i} = 0.4 \ell$ has volume fraction $n^\text{(f)} \simeq 0.5$ but negligible mass density $\rho^\text{(f)}$ and shear modulus $G^\text{(f)}$. Therefore, we have $\tilde G \simeq 0.33~G^\text{(s)}$ and $\bar \rho \simeq 0.5~\rho^\text{(s)}$, see Eqs.~\eqref{Effective} and \eqref{EffectiveRho}.
The maximum number of Fourier harmonics used in the FFT algorithm is $2 N_\text{max}+1$ where $N_\text{max} = 6$, and 90 wave vectors spaced linearly along the edges of the Brillouin zone are selected (i.e., 30 wave vectors per edge). In Fig.~\ref{fig:BG1}, the angular frequencies deduced from the Bloch-Floquet analysis are represented using solid lines, whereas the dispersion relationship deduced from the homogenised theory \eqref{EffectiveEq} is represented using dashed lines. The results obtained here are coherent with results by \citet{poulton00}, see also \citet{andrianov21}.

As expected, we observe that the leading order homogenisation theory \eqref{EffectiveEq} provides an accurate estimation of the dispersion relationship in the low frequency range. However, this theory is not able to capture the frequency band gap where no wave can propagate (grey rectangle in Fig.~\ref{fig:BG1}-bottom), which results from destructive wave interference. The wave dispersion effects leading to the emergence of the frequency band gap could be estimated based on high-order homogenisation theory \citep{andrianov21}.

Our calculations show also that the determinant of the matrix $[{\text D}_{\bm{G}'\bm{G}}]$ vanishes only at the exceptional point $\Gamma$, where the wave number $\kappa$ is equal to zero. Therefore, when waves propagate, invertion of Eq.~\eqref{Bloch}\textsubscript{2} shows that the magnetic potential $\tilde \phi$ in \eqref{BlochRep} is equal to zero. Therefore, there is no incremental magnetic induction associated with the propagation of antiplane shear waves in the periodic composite, in a similar fashion to the homogeneous material (Section~\ref{sec:SoL}), and consistently with the homogenised theory \eqref{EffectiveEq}. Finally, since there is no incremental magnetic field in the present case, we conclude that the TI magneto-elastic theory \eqref{EnergyTI} is still suitable to model the propagation of antiplane shear waves in a composite material that is subjected to a permanent magnetic field along the fibre direction. However, the present effective theory would possibly need to be revised if other loading types are considered.

\subsection{Band gap tunability}

Based on the above observations, we note that the uniaxial deformation along the fibre axis of stretch ratio $\lambda$ leads to the same modification of the effective shear modulus in the fibre and in the host medium. In both phases, the effective shear modulus equals $\tilde G \lambda^{\prime 2}$, where $\tilde G = G^\bullet$ and $\lambda = (\lambda')^{-2}$. However, the ratio $G^\text{(f)}/G^\text{(s)}$ between these effective shear moduli remains unchanged. In addition, the mass densities $\rho^\text{(f)}$ and $\rho^\text{(s)}$ are not modified by this uniaxial pre-deformation.

Let us recall that for a large slab of TI magneto-elastic material \eqref{EnergyTI} subjected to a static uniaxial motion along the fibre axis $\hat{\bf n}_0 = \hat{\bf e}_z$, the vertical stretch ratio can be deduced from the knowledge of the vertical normal force and of its magnetic part according to Eq.~\eqref{InterStretch}. In the case of a controlled normal force such that $\mathcal{F} = 1$, the stretch $\lambda$ is only function of the magnetic part $\mathcal{F}_h$ of the normal force, and we have the approximation \eqref{Taylor}, see also Fig.~\ref{fig:StretchMag}. 

Now, we go back to our example of composite material which we studied using the Bloch-Floquet method (Figure~\ref{fig:BG1}). Here, the band gap corresponds to the frequency interval $3.14529 < \omega \ell/c^\text{(s)} < 4.0484$ marked by a grey rectangle in the figure. Since $G^\text{(f)}/G^\text{(s)}$ and $\rho^\text{(f)}/\rho^\text{(s)}$ do not vary with the vertical stretch ratio, this dimensionless frequency interval remains unchanged when the material undergoes uniaxial stretching. However, we have $c^\text{(s)} = \lambda' \sqrt{G^\text{(s)} / \rho^\text{(s)}}$ and $\ell = \ell_0 \lambda'$, where $\ell_0$ denotes the size of the periodic unit cell in the undeformed material. Therefore, the ratio $\ell/c^\text{(s)}$ does not depend on the stretch, and the band gap for the angular frequency $\omega$ is unaffected by an applied tensile load of mechanical or magnetic type.

This result is reminiscent of results reported by \citet{zhang17}. It is also consistent with the results by \citet{galich17}, who found that a modification of stiffness ratios and of volume fractions has a sensible effect on the material's dispersion properties, while pre-stretch mostly influences the wave directivity patterns (Section~\ref{subsec:Tensile}). In contrast, using a hard-magnetic material model, \citet{padmanabhan24} found that the location and width of the band gaps depends on the magnitude and orientation of the applied magnetic field, and that ``the residual magnetic flux density has a positive influence on the tunability of band gaps''. We emphasise that this effect is not included in our study, given that soft magneto-active material models do not exhibit residual magnetic fluxes.

In the context of electro-active Gent composites, \citet{jandron18} found that ``the electrical tunability of band gaps is minimal'' prior the emergence of an electro-mechanical instability, when the material is subjected to electrical preloading parallel to the fibres. Moreover, they report that band gaps cannot be tuned in neo-Hookean composites subjected to a homogeneous state of deformation. A similar observation is made by \citet{bertoldi08} in the case of inert neo-Hookean and Gent elastomers subjected to finite deformations, where the effect of micro- and macro-instabilities on band gaps is also investigated. For further works dedicated to the study of patterns caused by instability in magneto-active composites, the interested reader is referred to \citet{goshkoderia20} as well as \citet{arora24}, where a soft inert matrix filled with rigid magnetic inclusions is considered. The buckling instability that develops in magneto-active laminates was investigated by \citet{chen23}.

\section{Conclusion}\label{sec:Conclu}

Let us summarise our results. We have derived an exact homogenisation result for a composite cylinder made of neo-Hookean magneto-elastic materials and subjected to out-of-plane shear with a permanent vertical magnetic field (see Appendix~\ref{app:I1} for a direct extension of our result to materials with generalised neo-Hookean fibres). The related effective theory \eqref{EnergyTI} can be used to model the propagation of shear waves in a fibre-reinforced composite material provided that it is subjected to a permanent magnetic field aligned with the fibres. This result was obtained by combining a small-on-large theory with periodic homogenisation and with the Bloch-Floquet technique.

Using the small-on-large theory, we have shown that in general, the state of total stress in the material cannot be inferred based on two shear wave speed measurements, unless other information is known. This specificity of magneto-elastic active materials contrasts with the case of passive material types, such as elastic and viscoelastic ones \citep{berjamin24}. Then, we have estimated the speed of shear waves propagating in such a homogeneous material subjected to uniaxial tension-compression, as well as the condition for its macroscopic stability. The relationship between the normal force and the amount of stretch in the material allows to induce a uniaxial tensile motion based on the application of a permanent magnetic field, and therefore to tune the speed of incremental shear waves. Based on the above results, we note that the wave directivity patterns in fibre-reinforced composites may vary with the magnitude of an applied magnetic field, while the frequency band gaps are unchanged. Similarly to the magnetically inert case, the effective response of fibre-reinforced media contrasts significantly with that of laminates in this regard \citep{galich17b,karami19}.

Potentially, a more general theory for the modelling of fibre-reinforced magneto-elastic media could be obtained based on more sophisticated homogenisation schemes. In fact, a more suitable effective theory is expected to involve further scalar invariants in the expression of the magneto-elastic energy function, see Eq.~\eqref{InvarExtra}. Furthermore, as shown in our dispersion analysis, the small-on-large homogenisation theory \eqref{EffectiveEq} could be extended to include higher-order terms of the small parameter $\epsilon$. In general, this process leads to effective partial differential equations with higher-order partial derivatives, and provides an accurate estimation of the dispersion band gaps \citep{wautier15,andrianov21}.

In its current form, the TI magneto-elastic theory \eqref{EnergyTI} needs to be revised for fibre-reinforced media subjected to other loading types than magnetic fields along the fibres, or for materials with a more complex magneto-active response. Here, the contribution of the magnetic field to the energy is a simple quadratic term (the square norm of the magnetic induction), which does not account for material anisotropy, nor for the magnetic saturation effect observed in ferromagnetic materials. Besides the restriction to magnetic fields aligned with the fibres, other limitations of this theory are the usual assumptions of homogenisation theory (long waves, no micro-instability). Future works might be dedicated to the study of high-contrast composites where local resonances can develop \citep{comi20}. Furthermore, the cases of other periodic fibre arrangements and of random micro-structures are also of interest, see for instance the works of \citet{galich18} as well as \citet{lopezpamies10}.

\subsection*{Acknowledgements}

HB is grateful to R\'emi Cornaggia (Sorbonne University, France) for insightful discussions. This project has received funding from the European Research Council (ERC) through Grant No. 852281 -- MAGIC.


\appendix

\section{Beyond neo-Hookean elasticity}\label{app:OtherConst}

\subsection{The Yeoh generalised neo-Hookean fibre}\label{app:I1}

We revisit the out-of-plane shearing problem \eqref{ShearComp} for the case of a generalised neo-Hookean fibre inside a neo-Hookean shell, see \citet{shmuel10} for the full generalisation. Thus, the magneto-elastic energy \eqref{EnergyTI} with the parameters $\tilde G = \bar G = G^\text{(s)}$ and $\mu^\text{(s)}$ for the shell material is unchanged, whereas the fibre is now described by the energy function
\begin{equation}
	\Psi = \frac{G^\text{(f)}}2 \left((I_1^\text{(f)} - 3) + \tfrac12 \varepsilon (I_1^\text{(f)} - 3)^2\right) + \frac{I_5^\text{(f)}}{2\mu^\text{(f)}} ,
	\label{Yeohfibre}
\end{equation}
together with its derivatives
\begin{equation}
	\Psi_1 = \frac{G^\text{(f)}}2 \left(1 + \varepsilon (I_1^\text{(f)} - 3)\right), \quad \Psi_5 = \frac{1}{2\mu^\text{(f)}} , \quad \Psi_7 = 0 .
\end{equation}
The parameter of nonlinearity $\varepsilon > 0$ is assumed small, and the neo-Hookean fibre is recovered for $\varepsilon\to 0$. The expression of the invariants is deduced from Eq.~\eqref{InvarComp}.

For such a composite material in equilibrium \eqref{Equil}, the Lagrangian multiplier $p^\bullet$ must be constant in each phase, and the functions $f^\bullet$ must still be of the form \eqref{deBottonF}. However, the coefficients $\alpha^\bullet$, $\beta^\bullet$ do no longer satisfy Eqs.~\eqref{deBotton}-\eqref{deBotton2}, and the expression of the pressure jump $p^\text{(f)} - p^\text{(s)}$ must be updated. Here, we seek approximate expressions in the form of truncated power series valid for small values of $\varepsilon$, i.e., we write $\alpha^\bullet \simeq \alpha^{\bullet}_0 + \varepsilon \alpha^{\bullet}_1$, and similar expansions are introduced for $\beta^\bullet$ and $p^\bullet$.

Upon enforcing the boundary conditions at the inner and outer interfaces of the composite cylinder depicted in Fig.~\ref{fig:Struct}, we recover the neo-Hookean case at leading order of the parameter $\varepsilon$, that is, $\alpha^\bullet_0$ and $\beta^\bullet_0$ are given by the expressions in Eqs.~\eqref{deBotton}-\eqref{deBotton2}, and $p^\text{(f)}_0 - p^\text{(s)}_0 = (G^\text{(f)} - G^\text{(s)}) \lambda^{\prime 2}$. The first-order corrections to these coefficients and to the pressure jump across the layers read
\begin{equation}
	\begin{aligned}
		&\alpha^\text{(f)}_1 = - \frac{n^\text{(s)}}{n^\text{(f)}} \alpha^\text{(s)}_1 , \quad 
    	\alpha^\text{(s)}_1 = \frac{n^\text{(f)} G^\text{(f)} \alpha^\text{(f)}_0 ([\bar I_1^\text{(f)}]_0-3)}{n^\text{(s)}G^\text{(f)} + (1+n^\text{(f)})G^\text{(s)}} , \\
    	&\beta^\text{(f)}_1 = 0 ,\quad
    \beta^\text{(s)}_1 = -\alpha^\text{(s)}_1 ,\\
    	&p^\text{(f)}_1 - p^\text{(s)}_1 = G^\text{(f)} \lambda^{\prime 2} ([\bar I_1^\text{(f)}]_0-3) ,
	\end{aligned}
	\label{deBottonCorr}
\end{equation}
where $[\bar I_1^\text{(f)}]_0 = \lambda^2 + 2\lambda^{\prime 2} + (\alpha^\text{(f)}_0 \gamma)^2$. We can now evaluate the leading order terms of the tractions applied to the lateral boundary, which are equal to
\begin{equation}
	\begin{aligned}
    &\sigma_{rr} = \left(\bar G + \varepsilon n^\text{(f)} G^\text{(f)} ([\bar I_1^\text{(f)}]_0 - 3) \right) \lambda^{\prime 2} - \bar p ,\\
    &\sigma_{r z} = \left(\tilde{G} + 2 \varepsilon \alpha^\text{(s)}_1 G^\text{(s)}\right)\lambda' \gamma  \cos\theta ,
    \end{aligned}
    \label{TractionsCorr}
\end{equation}
with the effective quantities \eqref{Effective} and $\bar{p} = n^\text{(f)} p^\text{(f)} + n^\text{(s)} p^\text{(s)}$. Thus, the magneto-mechanical response of the composite cylinder with Yeoh fibre deviates from the neo-Hookean response \eqref{EffTractions} unless $\varepsilon = 0$.

Now, let us evaluate the average magneto-elastic energy in the composite, $\bar \Psi = \sum_\bullet n^\bullet \bar\Psi^\bullet$. The phase-related average energy $\bar\Psi^\bullet$ is deduced from the energy function \eqref{Yeohfibre} for the fibre by spatial averaging, whereas the energy in the shell material is given by \eqref{EnergyTI}, with the single shear modulus $G^\text{(s)}$, magnetic permeability $\mu^\text{(s)}$, and the phase-related average invariants \eqref{InvarAver}. By using the asymptotic expansions of $\alpha^\bullet$ and $\beta^\bullet$ with respect to the small parameter $\varepsilon$, spatial averaging yields
\begin{equation}
	\begin{aligned}
	\bar \Psi &= \frac{\bar G}{2} (\lambda^2 + 2 \lambda^{\prime 2} - 3) + \frac{\tilde G}{2} \gamma^2 + \frac{\bar b^2}{2 \bar \mu} + \frac{\varepsilon n^\text{(f)} G^\text{(f)}}{4} ([\bar I_1^\text{(f)}]_0 - 3)^2 \\
	&\quad + \varepsilon \left(n^\text{(f)} G^\text{(f)} \alpha^\text{(f)}_0 \alpha^\text{(f)}_1 + n^\text{(s)} G^\text{(s)} (\alpha^\text{(s)}_0 \alpha^\text{(s)}_1 + \beta^\text{(s)}_0 \beta^\text{(s)}_1/n^\text{(f)})\right) \gamma^2 ,
	\end{aligned}
\end{equation}
at leading order in $\varepsilon$, where use was made of the identities \eqref{Effective}-\eqref{Effective2} satisfied by the coefficients $\alpha^\bullet_0$ and $\beta^\bullet_0$. Given the formulas \eqref{Effective}, \eqref{deBotton}, \eqref{deBotton2}, \eqref{deBottonCorr} and the expression of the invariants \eqref{InvarShear} for the homogeneous cylinder, this expression of the average energy is found coherent with an effective magneto-elastic energy function of the form
\begin{equation}
	\begin{aligned}
	\Psi &=  \frac{\tilde G}{2} (I_1 - 3) + \frac{\bar G - \tilde G}{2} \left(I_7 + 2 I_7^{-1/2} - 3\right) + \frac{\varepsilon n^\text{(f)} G^\text{(f)}}{4} Q^2 + \frac{I_5}{2 \bar \mu} , \\
	Q &= (\alpha^\text{(f)}_0)^2 (I_1-3) + (1-(\alpha^\text{(f)}_0)^2) \left(I_7 + 2 I_7^{-1/2} - 3\right) ,
	\end{aligned}
	\label{EnergyYeohTI}
\end{equation}
where the expression of $\alpha^\text{(f)}_0$ is provided in Eq.~\eqref{deBotton}. Together with
\begin{equation}
	\begin{aligned}
	\Psi_1 &= \frac{\tilde G}{2} + \frac{\varepsilon n^\text{(f)} G^\text{(f)}}{2} (\alpha^\text{(f)}_0)^2 Q ,\\
	\Psi_7 &= \left( \frac{\bar G - \tilde G}{2} + \frac{\varepsilon n^\text{(f)} G^\text{(f)}}{2} (1-(\alpha^\text{(f)}_0)^2) Q \right) \left(1 - I_7^{-3/2}\right) ,
	\end{aligned}
\end{equation}
and $\Psi_5 = 1/(2 \bar \mu)$, this result generalises Eqs.~\eqref{EnergyTI}-\eqref{Psi7Coeff}. Upon using Eqs.~\eqref{deBotton}-\eqref{deBottonCorr} and introducing the alternative Lagrangian multiplier $q$ such that
\begin{equation}
    p = q - \left(\bar G - \tilde G + \varepsilon n^\text{(f)} G^\text{(f)} (1-(\alpha^\text{(f)}_0)^2) Q\right) I_7^{-1/2} ,
\end{equation}
instead of \eqref{SubsBotton}, one can show that the tractions at the boundary of a homogeneous cylinder described by the energy function \eqref{EnergyYeohTI} are identical to the tractions \eqref{TractionsCorr} at the boundary of the heterogeneous cylinder. This completes the proof of our homogenisation result, which is valid for small values of the nonlinearity parameter $\varepsilon$.

\subsection{The second strain invariant}\label{app:I2}

Let us reconsider the out-of-plane shearing motion \eqref{Shear}. Here, we explore some consequences related to the substitution of the neo-Hookean strain energy $\frac12 \tilde{G} (I_1-3)$ with a similar one based solely on the second strain invariant $I_2$ instead of the invariant $I_1$ introduced in Eq.~\eqref{Invar}. Namely, we consider a strain energy function of the form $\frac12 \tilde{G} (I_2-3)$ where \citep{kuhl24}
\begin{equation}
    I_2 = \tfrac12 \big( I_1^2 - \text{tr} (\bm{C}^2) \big) = \lambda^{\prime 4} + 2\lambda + \lambda^{\prime 2}\gamma^2 .
\end{equation}
For such a material behaviour, the constitutive law in Eq.~\eqref{ConstitutiveExpr} needs to be modified in such a way that the term $\tilde{G}\bm B$ is replaced by $-\tilde{G}\bm{B}^{-1}$.

Here, we neglect the material anisotropy and the magneto-elastic coupling. In mechanical equilibrium, the tractions on the lateral boundary read
\begin{equation}
    \begin{aligned}
    \sigma_{rr} &= -\tilde{G}\lambda^{\prime 2} \gamma^2\cos^2\!\theta - \tilde{G}\lambda - p , \\
    \sigma_{r\theta} &= \tilde{G} \lambda^{\prime 2} \gamma^2\cos\theta\sin\theta , \quad
    \sigma_{rz} = \tilde{G} \lambda^{\prime 3} \gamma \cos\theta ,
    \end{aligned}
\end{equation}
where the Lagrangian mutiplier $p$ is a constant. Contrary to the neo-Hookean case \eqref{TShear}, we observe that the out-of-plane shear motion induces a $\theta$-dependent normal traction $\sigma_{rr}$ along the lateral surfaces of the cylinder, as well as a non-zero angular traction $\sigma_{r\theta}$.

The out-of-plane shear and extension problem is more complex for a composite material whose phases are isotropic and elastic, when the strain energy is a linear function of the second strain invariant $I_2^\bullet$. Based on similar assumptions as in the neo-Hookean magneto-elastic case \eqref{ShearComp}, we write the equilibrium equations in the reference configuration, and seek the pressure $p^\bullet$ as a function of $(R,\Theta)$ only. Unfortunately, it seems difficult to arrive at a general solution in the present case. Therefore, the homogenisation method used by \citet{debotton06b} does not seem straightforwardly applicable for the present constitutive law. Perhaps the major difficulty stemming from such a constitutive law is of algebraic nature, given that the resolution of the equilibrium equations is less tractable than for neo-Hookean models.

\section{Torsion and extension of a magneto-elastic composite cylinder}\label{app:Torsion}

In a similar fashion to the out-of-plane shearing motion studied earlier, we now consider the torsion and extension problem for a magneto-elastic cylinder. First, we tackle the case of a transversely isotropic homogenous cylinder described by the energy function \eqref{EnergyTI}. Then, the results obtained in the monophasic case are connected to those obtained in the case of a composite cylinder.

\subsection{The homogeneous TI cylinder}

We follow Section 4.2 of \cite{dorfmann04}. The cylinder is subjected to torsion and extension, which is defined by
\begin{equation}
    r = \lambda' R, \quad \theta = \Theta + \lambda \tau Z, \quad z = \lambda Z ,
    \label{Torsion}
\end{equation}
where $\tau$ is the amount of twist. In a cylindrical coordinate system, the deformation gradient tensor reads
\begin{equation}
    \bm{F} = \begin{bmatrix}
        \lambda' & 0 & 0\\
        0 & \lambda' & \lambda\tau r\\
        0 & 0 & \lambda
    \end{bmatrix} ,
    \label{FTorsion}
\end{equation}
which has unit determinant, and the vector ${\bf n}$ has no radial component. Furthermore, $I_7 = \lambda^2 (1 + \tau^2 r^2)$.

The solution of the magnetostatic equations was discussed earlier. On using the mechanical equilibrium equations \eqref{Equil} in cylindrical coordinates, we arrive at
\begin{equation}
    \begin{aligned}
    \sigma_{rr} &= \sigma_{rr}|_{r=a} + \tfrac12 \bar G \lambda^2 \tau^2 (r^2 - a^2) \\
    &\quad  + (\bar{G}-\tilde{G}) \big( I_7^{-1/2} - I_7|_{r=a}^{-1/2}\big)  ,
    \end{aligned}
    \label{PTorsion}
\end{equation}
which determines also $p = \tilde G \lambda^{\prime 2} - \sigma_{rr}$.

We can now evaluate the non-zero tractions applied to the cross section,
\begin{equation}
     \begin{aligned}
     \sigma_{\theta z} &= \big(\bar G + (\tilde G - \bar G) I_7^{-3/2} \big) \lambda^2\tau r ,\\
     \sigma_{z z} &= \big(\bar G + (\tilde G - \bar G) I_7^{-3/2} \big) \lambda^2 + hb - p .
     \end{aligned}
     \label{TTorsion}
\end{equation}
While the equilibrium equations are not modified by the application of a uniform and permanent vertical magnetic field, the latter modifies the value of the vertical traction. The computation of the torque $\mathcal{T}$ and normal force \eqref{NormalForce} follows. We find
\begin{equation}
    \mathcal{T} = 2\pi \int_0^a r^2 \sigma_{\theta z} \text{d}r \simeq \tfrac12 \pi A^4 \big(\bar G + (\tilde G - \bar G)\lambda^{-3}\big) \tau ,
    \label{Torque}
\end{equation}
and $\mathcal{N} \simeq \mathcal{A} - \mathcal{B}\tau^2$, with
\begin{equation}
    \begin{aligned}
    \mathcal{A} &= \pi A^2\bar{G} (\lambda - \lambda^{-2}) + \pi A^2 \big(hb + \sigma_{rr}|_{r=a}\big)/\lambda , \\
    \mathcal{B} &= \tfrac14 \pi A^4 \big(\bar G + 4 (\tilde G - \bar G)\lambda^{-3}\big) ,
    \end{aligned}
    \label{Poynting}
\end{equation}
where we have assumed $\tau A \ll 1$ for the sake of keeping mathematical expressions concise.

With the above expressions, it becomes apparent that a negative or reverse Poynting effect such that the normal force $\mathcal{N} \geq 0$ is tensile is potentially achievable for small amounts of twist. In fact, for a given stretch ratio $\lambda$, it then suffices to chose $\mathcal{A}$ positive and large enough, for instance by imposing a magnetic field of large amplitude. Alternatively, based on known results obtained for the standard reinforcing model \citep{horgan12}, a large stretch ratio $\lambda$, or a large degree of anisotropy $E = \frac32 (\bar G - \tilde G)$ are likely to produce a reverse Poynting effect as well.

\subsection{The composite cylinder}

We study a similar problem as in \citet{monteiro11}, where it is suggested that the simple torsion deformation with elongation is admissible in composites. A related problem was considered by \citet{wu15} for a cylinder of finite length.
Here, the cylinder is subjected to extension and torsion \eqref{Torsion} in each phase, with the same amount of twist $\tau$. Thus, the deformation gradient tensor $\bm{F}^\bullet$ for each phase is of the form \eqref{FTorsion}.

In equilibrium \eqref{Equil}, we can determine the expression of the pressure $p^\bullet(r^\bullet)$ in each phase based on the perfect bonding condition at the inner interface. Finally, upon integration of the moments and  tractions applied to the cross section, we obtain the following expression of the torque
\begin{equation}
    \mathcal{T} = \tfrac12 \pi A^4 \hat{G} \tau , \quad
    \hat{G} = (n^\text{(f)})^2 G^\text{(f)} + \big(1-(n^\text{(f)})^2\big) G^\text{(s)} .
\end{equation}
The normal force $\mathcal{N} = \mathcal{A} - \mathcal{B} \tau^2$ is deduced from
\begin{equation}
    \begin{aligned}
    \mathcal{A} &= \pi A^2\bar{G}(\lambda - \lambda^{-2}) + \pi A^2 \big(h\bar b + \sigma_{rr}|_{r=a}\big)/\lambda ,\\
    \mathcal{B} &= \tfrac14 \pi A^4 \hat{G} ,
    \end{aligned}
\end{equation}
and from the effective quantities defined earlier, see Eqs.~\eqref{Effective} and \eqref{EffectiveN}.

We observe that the composite cylinder's outer boundaries are subjected to the same torsional motion as the boundaries of the transversely isotropic cylinder, and that both cylinders are in equilibrium. Despite this equivalence between the two problems, a quick look at the transversely isotropic case \eqref{Torque}-\eqref{Poynting} reveals that in general, the related results are not strictly equivalent to those above due to distinct expressions for the coefficient $\mathcal{B}$, even in the limit of small amounts of twist $\tau A$. The observed inaccuracy relates to the fact that the pure torsion deformation is not macroscopically homogenous. Despite this drawback, the TI model \eqref{EnergyTI} still provides a satisfactory approximation in many cases \citep{debotton06b}.

\addcontentsline{toc}{section}{References}

\bibliography{cas-refs}{}

\end{document}